\documentclass[aps,prx,a4paper,twocolumn,superscriptaddress,longbibliography,floatfix]{revtex4-2}

\usepackage{amssymb,amsmath,amsfonts}
\usepackage{ esint }
\usepackage{graphicx}
\usepackage[dvipsnames]{xcolor}
\usepackage[bookmarksopen,colorlinks,linkcolor=blue,citecolor=olive,urlcolor=red]{hyperref}
\usepackage{bm}

\DeclareMathOperator{\re}{Re}
\DeclareMathOperator{\im}{Im}
\DeclareMathOperator{\tr}{tr}
\DeclareMathOperator{\Tr}{Tr}
\DeclareMathOperator{\arccosh}{arccosh}
\DeclareMathOperator{\arccot}{arccot}
\DeclareMathOperator{\sign}{sign}

\def\br{\mathbf{r}}
\def\bj{\mathbf{j}}

\def\bA{\mathbf{A}}
\def\ba{\mathbf{a}}

\def\eps{\varepsilon}

\newcommand{\corr}[1]{\langle#1\rangle}

\def\be{\begin{equation}}
\def\ee{\end{equation}}

\graphicspath{{pictures/}}

\begin{document}

\title{Optical conductivity of a dirty current-carrying superconductor}

\author{Artem V.\ Polkin}
\affiliation{L.~D.\ Landau Institute for Theoretical Physics,  Chernogolovka 142432, Russia}

\affiliation{Laboratory for Condensed Matter Physics, HSE University, Moscow 101000, Russia}

\author{Mikhail A.\ Skvortsov}
\affiliation{L.~D.\ Landau Institute for Theoretical Physics,  Chernogolovka 142432, Russia}

\affiliation{Moscow Institute of Physics and Technology, Dolgoprudny 141701, Russia}

\date{\today}

\begin{abstract}
We develop a full microscopic theory for the optical conductivity, $\sigma(\omega)$, of a dirty current-carrying superconductor. Within the Keldysh sigma model formalism, we obtain the general analytical expression for $\sigma(\omega)$, applicable for arbitrary frequency $\omega$, temperature $T$, and dc supercurrent $I$. 
In addition to altering the usual Mattis-Bardeen conductivity, $\sigma_1(\omega)$, a finite supercurrent introduces two new contributions: $\sigma_2^\text{qp}(\omega)$ from quasiparticle redistribution and $\sigma_2^\text{SH}(\omega)$ from the amplitude (Schmid-Higgs) mode excitation by the ac field. We investigate, both analytically and numerically, the main features of the optical conductivity in the presence of a dc supercurrent. They include a peak in $\re\sigma(\omega)$ above the optical gap and a sign change of $\im\sigma(\omega)$, with both effects becoming more pronounced at higher $I$ and lower $T$. We also elucidate the role of inelastic relaxation, which governs the low-frequency response, leading to a giant microwave absorption and a suppression of the apparent superfluid density at the critical current.
The optical conductivity measurement of a superconductor biased by a finite dc supercurrent enables the direct observation of the Schmid-Higgs mode via transport measurements.
\end{abstract}

\maketitle

\section{Introduction}

The electromagnetic response of superconductors has been the subject of extensive and ongoing research for decades \cite{MB, Wyatt1966, Dayem1967, Eliashberg1970, IvlevEliashberg1971, Eckern1979, Schmid1980, vanSon1984, Ovchinnikov1, Ovchinnikov2, Semenov2016, Crowley2022,TSK18}.
A fascinating manifestation of non-equilibrium physics is the microwave-driven enhancement of the superconducting gap discovered in Refs.\ \cite{Wyatt1966, Dayem1967} and theoretically explained by Eliashberg \cite{Eliashberg1970, IvlevEliashberg1971} as originating from the redistribution of quasiparticle population. 
In later works, Eliashberg's theory was generalized to include a finite supercurrent \cite{Schmid1980,vanSon1984} and arbitrary temperatures \cite{TSK18}.
Superconductivity enhancement arises in the second order in the amplitude of the microwave field as a correction to the period-average order parameter.

\begin{figure}[b]
\centering
\includegraphics[scale=0.7]{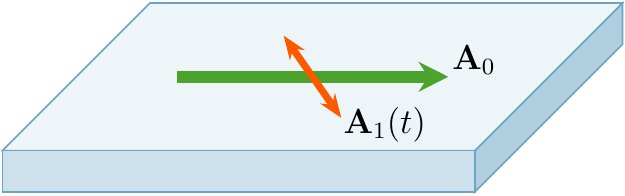}
\caption{A superconducting film carrying a dc supercurrent (with momentum $2e\mathbf{A}_0/\hbar c$) under an applied microwave field.}
\label{F:A+A}
\end{figure}

An even simpler, yet still very rich, phenomenon is the linear response of a superconductor, characterized by its frequency-dependent conductivity. With a finite dc supercurrent flowing through a superconductor, the optical conductivity becomes a tensor, which can be written as a sum of two contributions \cite{Ovchinnikov1, Ovchinnikov2, Moor2017}:
\be
\label{sigma}
  \sigma_{\alpha\beta}(\omega)
  =
  \sigma_1(\omega) \delta_{\alpha\beta}
  +
  \sigma_2(\omega) \hat A_\alpha \hat A_\beta .
\ee
Here $\hat{\mathbf A}$ is a unit vector in the direction of the superfluid momentum $2e\mathbf{A}_0/\hbar c$ (see Fig.\ \ref{F:A+A}), and the contribution $\sigma_2(\omega)$ vanishes at zero  supercurrent.

The optical conductivity of a BCS superconductor in the absence of a supercurrent, $\sigma_\text{MB}(\omega)\equiv\sigma_1(\omega)$ at $A=0$, was calculated by Mattis and Bardeen (MB) in 1958 \cite{MB}. 
Their theory, built on the Kubo formalism, generalizes the classical Drude result to dirty superconductors, with nonmagnetic impurity scattering serving as the source of momentum relaxation. The existence of the spectral gap $\Delta_0(T)$ results in the suppression of dissipative conductivity $\re\sigma_\text{MB}(\omega)$, at low temperatures, unless $\hbar\omega>2\Delta_0(T)$ and quasiparticles can be excited across the gap. At the same time, the reactive part of the optical conductivity diverges at low frequencies, $\omega\ll\Delta_0(T)$, as 
\be
\label{Im-sigma-SC}
  \im\sigma(\omega)/\sigma_0\approx\Omega_s/\omega ,
\ee
which is a hallmark of the condensate's superconducting response.
The frequency $\Omega_s$ proportional to the superfluid density is given by \cite{AG1959}
\be
\label{Omegas}
  \Omega_s(T) = \pi\Delta_0(T) \tanh[\Delta_0(T)/2T] .
\ee

The MB theory, originally developed for BCS superconductors, has been extended to include depairing effects from paramagnetic impurities \cite{Skalski1964}.
In this situation, the order parameter $\Delta(T)$ is suppressed compared to the BCS value $\Delta_0(T)$, the spectral gap is renormalized down to $E_g=(\Delta^{2/3}-\Gamma^{2/3})^{3/2}$, and the coherence peak is smeared at the energy scale $\omega_*\sim\Gamma^{2/3}\Delta^{1/3}$, where $\Gamma$ is the spin-flip rate \cite{AG1960}. 
As a result, the low-temperature optical gap shifts from $2\Delta_0(T)$ to $2E_g$, and the logarithmic divergence of $\re\sigma_\text{MB}(\omega)$ saturates at $\omega\lesssim\omega_*$.
Still, modifying the MB expression $\sigma_\text{MB}(\omega) \to \sigma_1(\omega)$ requires only adjusting the quasiparticle Green functions to account for depairing effects at a finite $\Gamma$ \cite{Fominov}. This term coincides with $\sigma_1(\omega)$ in Eq.\ \eqref{sigma}, provided $\Gamma$ is the depairing rate due to a finite supercurrent.

A key new feature brought by the moving superconducting condensate is the emergence of the second term in Eq.\ \eqref{sigma}, which is governed by the response functions of quasiparticles and the order parameter to an ac vector potential $\bA_1(\omega)$ \cite{Ovchinnikov1, Ovchinnikov2}:
\be
\label{sigma2-2}
  \sigma_2(\omega) 
  = 
  \sigma_2^\text{qp}(\omega) + \sigma_2^\text{SH}(\omega) .
\ee
In a dirty superconductor, $\sigma_2^\text{qp}(\omega)$ is expressed in terms of dynamic diffusons and cooperons, whereas $\sigma_2^\text{SH}(\omega)$ appears through the excitation of the amplitude mode of the order parameter.

Collective amplitude mode of the complex order parameter, which is now commonly referred as Schmid-Higgs (SH) mode, has been studied theoretically since 1968 \cite{Schmid1968, Littlewood1981, Littlewood1982}. However, its experimental evidence was long hindered by the weak coupling of the SH mode to electromagnetic field. In the 2010s, breakthroughs in terahertz spectroscopy enabled the first direct observations of the SH mode \cite{Matsunaga2012, Matsunaga2014, Beck2013, Sherman2015, Katsumi2020}. These experimental results triggered a resurgence of interest in the theoretical community \cite{Cea2015, Wang2025, Sun2020}, with the most prominent topics of research being dynamics of the homogeneous \cite{Derendorf2024} and spatially periodic \cite{Burmi2025} perturbation, and third harmonic generation, caused by SH mode \cite{Burmi2025,Derendorf2024,Silaev2019}, both of which were done also in presence of magnetic impurities \cite{Li2024,DzKamenev2025}.

At the same time, direct coupling of the SH mode to an external ac electric field in a current-carrying superconductor \cite{Ovchinnikov1, Ovchinnikov2, Moor2017} makes the optical conductivity a perfect tool for the observation of the SH mode, which, via the term $\sigma_2^\text{SH}(\omega)$, is responsible for the absorption peak above the optical gap, $2E_g$. Recently, the effect was theoretically studied in Refs.\ \cite{Jujo2022,Kubo2025}, where the zero-temperature limit was addressed.

\begin{figure*}
  \centering
  \includegraphics[width=\textwidth]{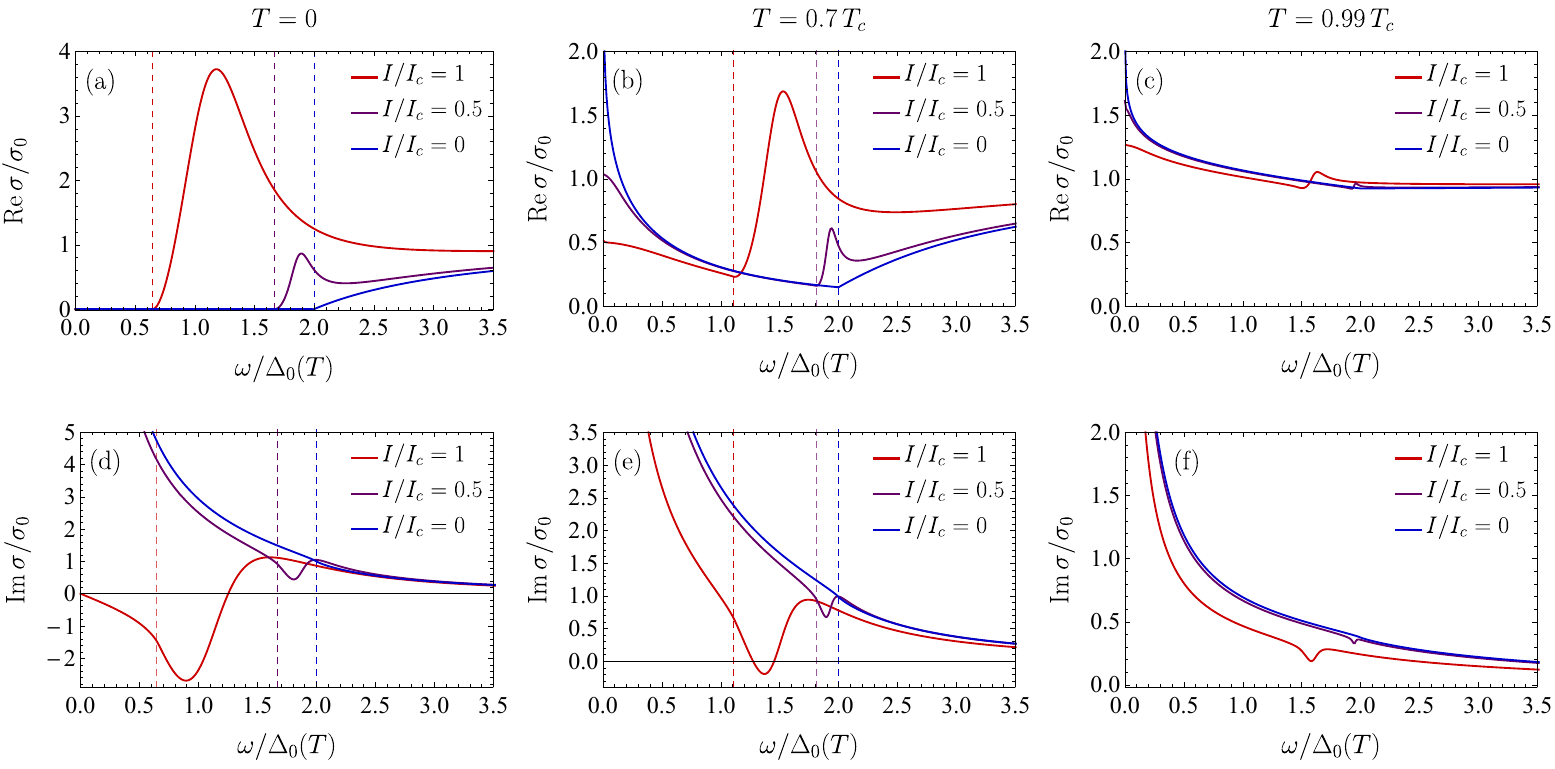}
\caption{Frequency dependence of the real (a, b, c) and imaginary (d, e, f) parts of the conductivity $\sigma(\omega)=\sigma_1(\omega)+\sigma_2(\omega)$ in the collinear geometry $\bA_1{\parallel}\bA_0$
at temperatures $T=0$ (a, d), $T=0.7\,T_c$ (b, e), and $T=0.99\,T_c$ (c, f). Three curves at each panel correspond to different dc supercurrents: $I/I_c(T)=0$, 0.5, and 1. Dashed lines mark the positions of $2E_g$ for given $T$ and~$I$.}
  \label{F:ReIm-plots}
\end{figure*}

A striking features of $\sigma_2(\omega)$ is the singular behavior of its quasiparticle contribution at small frequencies that should be regularized by the energy relaxation rate $\gamma$: $\sigma_2^\text{qp}(\omega)\propto1/(\omega+i\gamma)$. 
This expression derived in Ref.\ \cite{Ovchinnikov2} in the vicinity of the critical temperature, $T_c$, suggests that $\sigma_2(\omega)$ may even exceed $\sigma_1(\omega)$ at small frequencies, provided that $\gamma$ is sufficiently small. 
This growth of dissipation at low $\omega$ is reminiscent of the lower frequency bound $\omega>3.23\,\gamma$ for superconductivity stimulation by microwaves observed near $T_c$ \cite{TSK18}.

Recently, the problem of the microwave absorption in a superconductor with a finite dc supercurrent has been addressed in Ref.\ \cite{Spivak1}, where the contribution $\sigma_2^\text{qp}(\omega)$ has been interpreted in terms of Debye relaxation mechanism (this idea has been further elaborated in a number of subsequent publications \cite{Spivak2,Spivak3}). Near the critical temperatures, the phenomenological model of Ref.\ \cite{Spivak1} reproduces the Lorentz form of $\sigma_2^\text{qp}(\omega)$, but with the numerical coefficient differing from that obtained in Ref.\ \cite{Ovchinnikov2}. At the same time, the SH contribution, $\sigma_2^\text{SH}(\omega)$, was completely neglected in Ref.\ \cite{Spivak1}.

Motivated by unresolved theoretical inconsistencies---specifically, the disagreement between Refs.\ \cite{Spivak1} and \cite{Ovchinnikov2} near $T_c$, and between Refs.\ \cite{Kubo2025} and \cite{Ovchinnikov1} at low temperatures---we revisit the linear response of a current-carrying superconductor to microwave radiation at frequency $\omega$. Assuming the dirty limit ($T_c\tau\ll1$, where $\tau$ is the elastic scattering time), we calculate the optical conductivity, derive the general expressions for $\sigma_1(\omega)$ and $\sigma_2(\omega)$ in Eq.\ \eqref{sigma} at arbitrary $\omega$, $T$ and dc supercurrent $I$, and analyze them in various asymptotic regions. 
Using the language of the Keldysh nonlinear sigma model \cite{FLS2000}, we reproduce the findings of Ovchinnikov \emph{et~al.} \cite{Ovchinnikov1, Ovchinnikov2}, confirming them within a complementary theoretical framework.

The inelastic energy relaxation rate, coming into play at low frequencies through the contribution $\sigma_2(\omega)$, is related to thermalization. In the limit $\omega\to0$, the quasiparticle distribution adiabatically follows the applied field, maintaining local thermal equilibrium.
However, the crossover frequencies differ between $\sigma^\text{qp}_2(\omega)$ and $\sigma^\text{SH}_2(\omega)$. For the quasiparticle contribution, $\sigma^\text{qp}_2(\omega)$, it coincides with the single-particle inelastic relaxation rate $\gamma$. For the SH contribution, $\sigma^\text{SH}_2(\omega)$, it is given by
\be
\label{gammaQ-def}
  \gamma_Q \sim \sqrt{1-T/T_c} \, \gamma ,
\ee
which is known as the branch imbalance relaxation rate \cite{Tinkham1972,ArtemenkoVolkov1979,Hubler2010}. In the vicinity of $T_c$, $\gamma_Q \sim [\Delta(T)/\Delta(0)]\gamma \ll \gamma$.

As the smaller of the two rates $\gamma$ and $\gamma_Q$, the latter determines the crossover frequency between quasiequilibrium and nonequilibrium regimes. Correspondingly, the limit $\omega\ll\gamma_Q$ ($\omega\gg\gamma_Q$) will be referred to as \emph{quasistatic} (\emph{dynamic}).

Access to the general expression for the optical conductivity as a function of $\omega$, $T$ and $I$ enables us to map the evolution of key physical effects across the full parameter space of frequency, temperature, and supercurrent. They include
\begin{itemize}
\vspace{-1mm}
\item
peak in $\re\sigma(\omega)$, originating due to the SH mode excitation, which grows with decreasing $T$ and increasing $I$,
\vspace{-1mm}
\item
giant low-frequency dissipation determined by the inelastic relaxation, which however is exponentially suppressed at low temperatures,
\vspace{-1mm}
\item
formation of the region with negative $\im\sigma(\omega)$ at low temperatures and high currents,
\vspace{-1mm}
\item
sensitivity of the low-frequency inductive response, $\im\sigma(\omega)$, 
to the inelastic relaxation rate, with vanishing $\Omega_s$ at $I=I_c(T)$ for all $T$ in the quasistatic regime and only at $T=0$ in the dynamic regime.
\end{itemize}

The paper is organized as follows.
In Sec.\ \ref{S:Model&Results} we introduce the model and provide an extensive overview of numerical results, outlining the main trends in $\sigma(\omega)$. Section \ref{S:Tech} contains a step-by-step sigma-model derivation of the linear response at a finite dc supercurrent. General expressions for the three conductivity contributions, $\sigma_1$, $\sigma_2^\text{qp}$ and $\sigma_2^\text{SH}$, are presented in Sec.\ \ref{S:eqs-gen}. Their asymptotic behavior in the vicinity of $T_c$ is analyzed in Sec.\ \ref{S:Tc}.
The results are discussed in Sec.\ \ref{S:Discussion} and summarized in Sec.\ \ref{S:Summary}. 
Technical details of the calculations are relegated to several Appendices.

\section{Model and qualitative results}
\label{S:Model&Results}

\subsection{Model}

We consider a diffusive superconducting film or wire that is biased by a dc supercurrent and subjected to microwave irradiation. The absolute value of the order parameter is assumed to be uniform, and gauging out its phase one arrives at the vector potential
\be
\label{A(t)}
  \mathbf{A}(t) = \mathbf{A}_0 + \mathbf{A}_1 \cos\omega t ,
\ee
corresponding to the superfluid momentum $2e\mathbf{A}/\hbar c$.
The static component, $\bA_0$, is responsible for the supercurrent density $\bj = - Q(T,A_0) \bA_0$. The kernel $Q(T,A_0)$ is a nonlinear function of $A_0$, with $Q(T,0)=n_se^2/mc$ related to the superfluid density $n_s$. 
A flowing supercurrent acts as a source of depairing characterized by the rate
\be
\label{eqn:2:GammaDef}
  \Gamma = 2 D (e\mathbf{A}/\hbar c)^2 ,
\ee
where $D$ is the diffusion coefficient.

We study the linear response of a superconductor to the applied ac field $\bA_1$, which may have an arbitrary angle with respect to $\bA_0$, see Fig.\ \ref{F:A+A} (in the wire geometry, $\bA_1{\parallel}\bA_0$).
We derive the general expression for the conductivity tensor \eqref{sigma} valid for arbitrary $T<T_c$ and $j<j_c(T)$. The critical current density $j_c(T)$ is determined numerically by simultaneously solving the Usadel and self-consistency equations \cite{ClemKogan2012}.

Analytic formulae for $\sigma_1(\omega)$ and $\sigma_2(\omega)$ that yield the conductivity for an arbitrary angle between $\bA_1$ and $\bA_0$ are presented in Sec.\ \ref{S:eqs-gen}. Below we discuss the results in the collinear case $\bA_1{\parallel}\bA_0$, when the deviation of $\sigma(\omega) = \sigma_1(\omega)+\sigma_2(\omega)$ from the MB-like expression $\sigma_1(\omega)$ is most pronounced.

\subsection{Results overview}

In Fig.\ \ref{F:ReIm-plots} we visualize the frequency dependence of $\sigma(\omega)$, normalized by the normal-state Drude conductivity $\sigma_0$, in the parallel configuration $\bA_1\parallel\bA_0$. The real and imaginary parts are shown in the upper and lower panels, respectively, while different columns correspond to different temperatures: $T=0$ (left), $0.7\, T_c$ (middle), and $0.99\, T_c$ (right). The curves at each panel are obtained for various dc supercurrent: $I/I_c(T)=0$ (standard MB result), 0.5 and 1 (maximum possible contribution of $\sigma_2$). The frequency in Fig.\ \ref{F:ReIm-plots} is normalized by the equilibrium order parameter $\Delta_0(T)$ at zero supercurrent.

\begin{figure}
\centering
\includegraphics[width=0.98\linewidth]{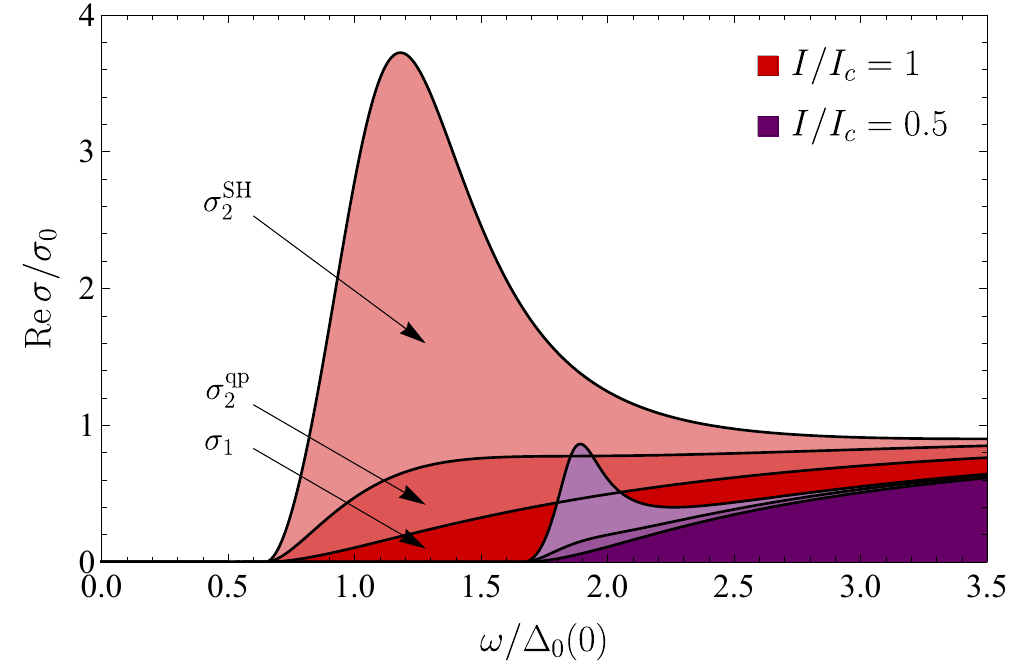}
\caption{Frequency dependence of $\re\sigma(\omega)$ for $\bA_1{\parallel}\bA_0$ at $T=0$, as shown in Fig.\ \ref{F:ReIm-plots}(a), with emphasized contributions of $\sigma_1(\omega)$, $\sigma_2^\text{qp}(\omega)$, and $\sigma_2^\text{SH}(\omega)$. 
Note that the absorption peak observed above the optical gap entirely originates from the excitation of the Schmid-Higgs mode.}
  \label{F:T0}
\end{figure}

\subsubsection{Dissipative conductivity}

We start with discussing the dissipative conductivity.

\emph{At zero temperature}, the MB conductivity $\re\sigma_\text{MB}(\omega)$ is strictly zero for $\omega<2\Delta_0(0)$ and monotonically increases for $\omega>2\Delta_0(0)$, approaching $\sigma_0$ at high frequencies. The presence of a finite supercurrent modifies $\re\sigma(\omega)$ in two ways: (i) the optical gap is reduced from $2\Delta_0(0)$ to $2E_g$ and (ii) a peak arises above the gap \cite{Ovchinnikov1,Ovchinnikov2, Jujo2022}. 
The peak's width and magnitude increase with the dc current, reaching their maximum values at the critical current. At $I=I_c(0)$, with the spectral gap suppressed down to $E_g=0.323\,\Delta_0(0)$ \cite{Maki1963,Skalski1964}, the width of the peak is comparable to $2E_g$ and its height is as large as $3.723\,\sigma_0$. 
At $I\ll I_c(0)$, the peak's height scales as $\sigma_\text{peak}/\sigma_0 \sim (\Gamma/\Delta)^{2/3} \ln^2(\Delta/\Gamma)$ \cite{Ovchinnikov2}.

Figure \ref{F:T0} replots $\re\sigma(\omega)$ at $T=0$ from Fig.\ \ref{F:ReIm-plots}(a), highlighting the relative contributions of its three components, $\sigma_1(\omega) + \sigma_2^\text{qp}(\omega) + \sigma_2^\text{SH}(\omega)$, by different fillings.
This figure clearly demonstrates that both $\sigma_2^{\mathrm{qp}}$ and $\sigma_2^{\mathrm{SH}}$ exhibit peaks, with the latter being substantially larger. Therefore, the peak in $\re\sigma(\omega)$ originates primarily from the amplitude Schmid-Higgs mode excitation. This result agrees with the findings of Ref.~\cite{Wang2025} in the dirty limit.

\textit{At finite temperatures}, the peak in the dissipative conductivity above the optical gap, appearing at finite $I$, persists. Its height decreases monotonically as the temperature increases, a trend directly linked to the suppression of $\Delta$ with rising $T$.

Yet a more striking difference between the $T=0$ and $T\neq0$ cases is the emergence of a finite $\re\sigma(\omega)$ below the optical gap, at $\omega < 2E_g$, see Figs.\ \ref{F:ReIm-plots}(b) and (c).
Originating from the heating of thermally excited quasiparticles, the sub-gap conductivity decreases exponentially with temperature: $\re\sigma(\omega)\propto e^{-\Delta(T)/T}$. Its frequency dependence for a given $T$ is more interesting. In the absence of a supercurrent, $\re\sigma_\text{MB}(\omega)$ exhibits a weak $\ln(\Delta/\omega)$ divergency produced by two nearly overlapping coherence peaks [see Eq.\ \eqref{sigma1-large-w}]. A nonzero current broadens the coherence peak, which acquires a width $\omega_*\sim\Gamma^{2/3}\Delta^{1/3}$, where $\Gamma$ is the associated depairing rate. Therefore, the $\ln(\Delta/\omega)$ divergency of $\re\sigma_1(\omega)$ at $\omega\gg\omega_*$ translates into its saturation at $\omega\ll\omega_*$ [see Eq.\ \eqref{sigma1-w<w*}] that can be clearly seen in panels (b) and (c).

\begin{figure}
\centering
\includegraphics[width=0.99\linewidth]{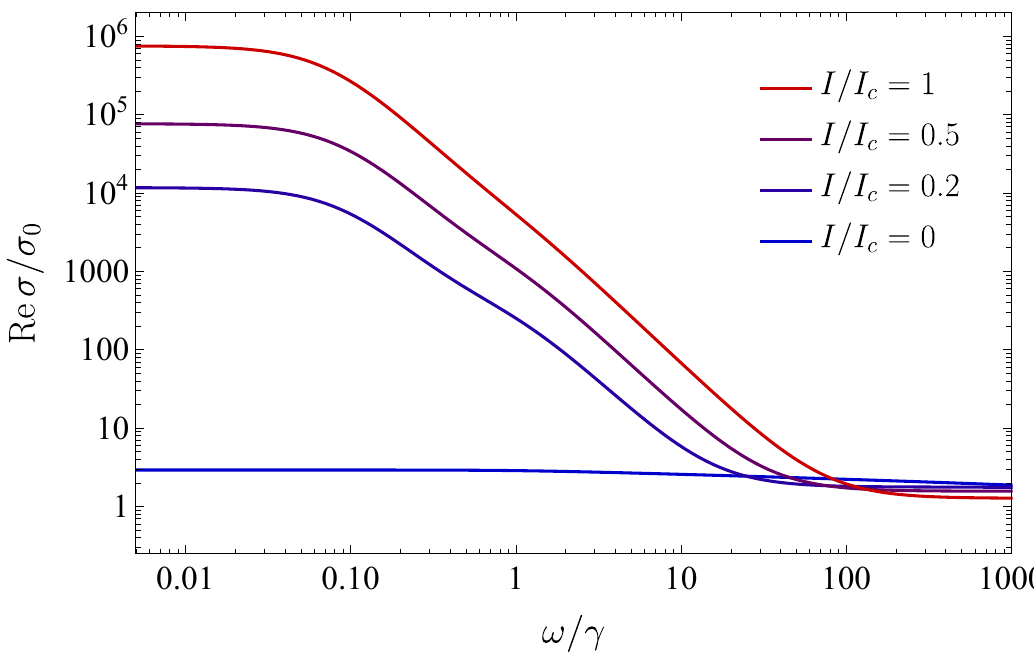}
\caption{Logarithmic plot of the low-frequency dissipative conductivity at $T=0.99\,T_c$ [Fig.\ \ref{F:ReIm-plots}(c)], computed for the inelastic rate $\gamma=10^{-6} \Delta_0(0)$ [$\Delta_0(0)$ is the gap at $T=0$ and $I=0$].}
  \label{F:low-w-log}
\end{figure}

Though the current-induced contribution $\re\sigma_2(\omega)$ is smaller than $\re\sigma_1(\omega)$ at $\omega\sim\omega_*$, its magnitude grows with decreasing frequency. Provided the inelastic relaxation rate is sufficiently small, at low frequencies $\re\sigma_2(\omega)$ may considerably exceed the bare Drude conductivity $\sigma_0$. This huge enhancement of dissipation at $T=0.99\,T_c$ is illustrated in Fig.\ \ref{F:low-w-log}, which shows the low-frequency behavior for $\gamma=10^{-6} \Delta_0(0)$ that cannot be resolved in Fig.\ \ref{F:ReIm-plots}(c). With $\re\sigma_2^\text{qp}(\omega)$ saturating at $\omega\sim\gamma$ and $\re\sigma_2^\text{SH}(\omega)$ saturating at $\omega\sim\gamma_Q\ll\gamma$, the frequency behavior of $\sigma(\omega)$ exhibits a number of crossovers determined by the competition between the quasiparticle and Schmid-Higgs contributions. The effect is maximal at the critical current. In the quasistatic limit ($\omega\ll\gamma_Q$) and at $I\sim I_c$, the enhancement determined by the Schmid-Higgs term is as large as $\re\sigma_2^\text{SH}(0)/\sigma_0\sim\Delta/\gamma$.

\begin{figure}
\centering
\includegraphics[width=\linewidth]{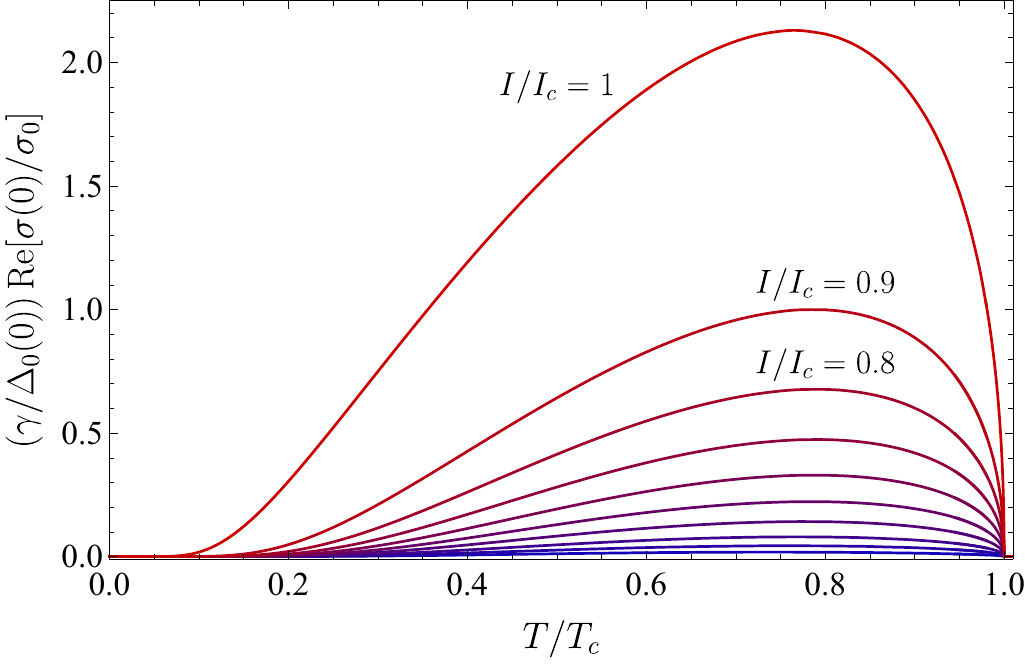}
\caption{Temperature dependence of the dissipative conductivity in the \emph{quasistatic} limit, $\re\sigma(0)\propto1/\gamma$, computed for $I/I_c(T)=0.1$, 0.2, $\dots$, 1, and the temperature-independent inelastic relaxation rate $\gamma = 10^{-6}\Delta_0(0)$. 
In this case, the giant absorption is exponentially suppressed at low temperatures due to the freezing of quasiparticles.}
\label{fig:reSigmaAtZeroOmega}
\end{figure}

In the Ginzburg-Landau region near $T_c$, $\re\sigma(0)/\sigma_0$ increases with decreasing temperature, driven by the growth of $\Delta(T)$, see Fig.~\ref{fig:reSigmaAtZeroOmega}.
However, the freezing out of thermal quasiparticles leads to an exponential suppression of the dissipative conductivity  at low temperatures: $\re\sigma_2(0)\propto e^{-E_g/T}/\gamma$.
For a temperature-independent inelastic relaxation rate, the maximal absorption enhancement occurs at $T\approx0.8\,T_c$, as shown in Fig.\ \ref{fig:reSigmaAtZeroOmega}.

The model of constant $\gamma(T)$ describes tunneling to a normal reservoir, see Sec.\ \ref{S:Tech}. 
For scattering mechanisms characterized by a power-law $\gamma(T)$, such as electron-electron or electron-phonon scattering, the low-temperature absorption still remains exponentially suppressed. 

Relaxation via two-particle recombination constitutes a special case, as its rate $\gamma(T)\propto e^{-E_g/T}$ is itself exponentially suppressed. 
The cancellation of the leading exponent in this case was noted in Ref.\ \cite{Spivak1}, which provided the estimate $\re\sigma_2(0)\propto T^{-1/2}$. In contrast, our analysis demonstrates that $\lim_{T\to0}\re\sigma_2(0)=0$. However, this point is largely academic, as the quasistatic limit, which requires exponentially low frequencies, is practically unattainable for recombination relaxation.

\subsubsection{Imaginary part of the conductivity \label{sec:imSigmaGen}}

Now we turn to the analysis of $\im\sigma(\omega)$ presented in the bottom raw of Fig.~\ref{F:ReIm-plots}. In the absence of a supercurrent, the MB theory predicts a monotonic growth of $\im\sigma_\text{MB}(\omega)$ with decreasing frequency, qualitatively similar for all temperatures. In the limit $\omega\ll\Delta$, the condensate motion results in a purely inductive response $\im\sigma_\text{MB}(\omega)=\Omega_s(T)/\omega$ [see Eq.\ \eqref{Im-sigma-SC}], where $\Omega_s(T)$ proportional to the superfluid density is given by Eq.\ \eqref{Omegas}.

A new feature introduced by a finite supercurrent is the appearance of the dip at $\omega \sim 2E_g$ originating from the term $\sigma_2(\omega)$. As a counterpart of the peak in $\re\sigma(\omega)$, the dip in $\im\sigma(\omega)$ is also enhanced with increasing $I$ and decreasing $T$.
Remarkably, the value at the dip, $\min\im\sigma(\omega)$ becomes negative at sufficiently low $T$ and sufficiently high $I$. According to Fig.\ \ref{F:ReIm-plots}(e), the region of negative $\im\sigma(\omega)$ at $I=I_c$ exists already at $T=0.7\,T_c$. At low temperatures, this region broadens, as evidenced in panel (d). Appearance of the negative $\im\sigma(\omega)$ region was recently brought up in Ref.~\cite{Wang2025}.

Figure \ref{F:ReIm-plots}(d) also reveals that at $T=0$ and $I=I_c$, $\im\sigma(\omega)$ does not diverge but drops to zero at $\omega=0$. This peculiar behavior results from a precise cancellation between the $1/\omega$ terms 
in $\sigma_1(\omega)$, $\sigma_2^\text{qp}(\omega)$, and $\sigma_2^\text{SH}(\omega)$, leading to the vanishing of $\Omega_s$ at zero temperature and the critical current (see Appendix \ref{A:SF0}).

\begin{figure}
\centering
\includegraphics[width=0.99\linewidth]{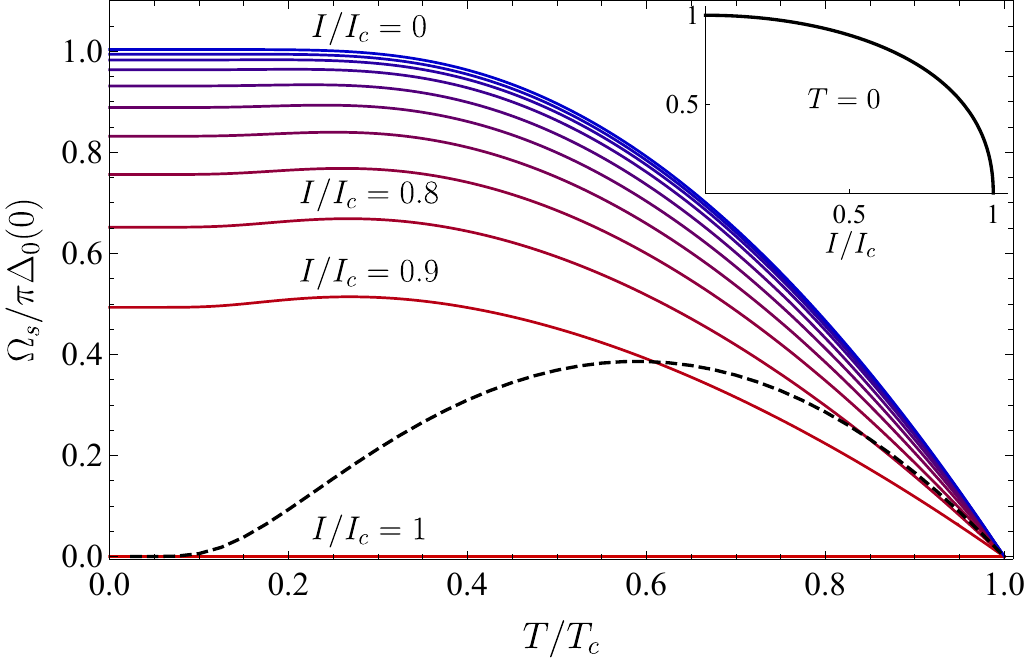}
\caption{Temperature dependence of $\Omega_s$, which governs the low-frequency superfluid inductive response $\im\sigma(\omega)/\sigma_0=\Omega_s/\omega$, computed in the \emph{quasistatic} limit, for $I/I_c(T)=0$, 0.1, $\dots$, 1. At zero current, $\Omega_s(T)$ follows Eq.\ \eqref{Omegas}. $\Omega_s$ is suppressed at finite currents $I$ and vanishes completely at the critical current $I_c(T)$ for all temperatures. Inset demonstrates $\Omega_s$ vs.\ $I/I_c(0)$ at $T=0$ [see Eq.\ \eqref{OmegasT0qst}]. The dashed line shows $\Omega_s$ at $I=I_c(T)$ computed in the \emph{dynamic} limit, $\omega\gg\gamma_Q$, but for $\omega\ll\omega_*$.}
\label{fig:coeffAAtZeroOmega}
\end{figure}

Since the conductivity contribution $\sigma_2(\omega)$ is sensitive to the inelastic relaxation rate $\gamma$, which is assumed to be small, the limit $\omega\to0$ should be taken with care. The results presented in Fig.\ \ref{F:ReIm-plots} are obtained in the \emph{dynamic} limit, $\omega\gg\gamma_Q$. The low-frequency behavior in the \emph{quasistatic} limit, $\omega\ll\gamma_Q$, is essentially different, with Fig.~\ref{fig:coeffAAtZeroOmega} demonstrating $\Omega_s$ as a function of $T$ and $I$. We see that in the quasistatic limit, the superconducting inductive response is lost, and $\Omega_s=0$ at the critical current for all temperatures.
This can be rationalized by recognizing that the limit $\omega\to0$ for an ac drive is equivalent to applying an infinitesimal dc current. Since the system is already at $I_c$, any additional current will destroy superconductivity.

To illustrate the difference between the dynamic and quasistatic regimes, we plot $\Omega_s(T)$ at $I=I_c$ obtained numerically in the dynamic regime by the dashed line in Fig.\ \ref{fig:coeffAAtZeroOmega}.
In this case, $\Omega_s$ vanishes only at $T=0$, in accordance with panels (d)--(f) of Fig.\ \ref{F:ReIm-plots}.

\section{Technicalities}
\label{S:Tech}

\subsection{Sigma-model description}

Non-equilibrium effects in dirty superconductors are described in terms of the semiclassical Keldysh Green's function $\check{G}$, which obeys the dynamic Usadel equations derived in Ref.\ \cite{LO-noneq}. However, analyzing the resulting system of equations, augmented by the self-consistency condition for the order parameter, remains challenging even for a small irradiation amplitude. This approach was adopted by Ovchinnikov and Isaakyan \cite{Ovchinnikov2}. Their work followed Ovchinnikov's earlier analysis in the Matsubara representation that required analytic continuation~\cite{Ovchinnikov1}. To verify their results and elucidate the details of the derivation, we will employ the Keldysh nonlinear sigma-model formalism \cite{FLS2000}, leveraging its powerful analytical tools for systematic perturbative expansion in terms of diffusons, cooperons and fluctuation propagator \cite{Fin,Efetov-book}. Though both approaches are equivalent, calculations in the sigma-model language are more elegant and straightforward.

In the uniform situation considered, the sigma model is a field theory for the matter field $Q_{t,t'}$ and the order parameter fields $\Delta(t)$, $\Delta_\text{q}(t)$.
In the saddle-point approximation, $Q$ coincides with the nonequilibrium semiclassical Green's function $\check{G}$. The field $Q_{t,t'}$, as well as its Fourier transform to the energy domain $Q_{\eps,\eps'}$, is a matrix in the Keldysh (K) and Nambu (N) spaces (Pauli matrices $\sigma_i$ and $\tau_i$, respectively). 
The sigma-model action $S$, which determines the weight $e^{i S}$ in the functional integral, is given by \cite{FLS2000,TSK18}
\be
\label{eqn:3:usadel}
  S 
  = 
  \frac{\pi i}{\delta} 
  \Tr\left[
	\Sigma
	Q - \frac{D}{2} (\check{\mathbf{a}} \tau_3 Q)^2 
  \right]
  - 
  \frac{4}{\lambda\delta}
  \Tr \Delta\Delta_\text{q} .
\ee
Here Tr stands for the matrix trace in the $\mathrm{K}\otimes \mathrm{N}$ space and also involves integration over energies (which hereafter will be treated as an additional continuous matrix index),
$\delta=1/\nu V$ is the mean level spacing (expressed via the density of states at the Fermi level $\nu$ and the system's volume $V$), $\lambda$ is the superconducting coupling constant, and 
\begin{equation}
\label{Sigma-def}
    \Sigma 
    = 
    i \eps \tau_3 - \check{\Delta} \tau_1 
    - (\gamma/2) Q_\text{res}.
\end{equation}
The last term in Eq.\ \eqref{Sigma-def} mimics inelastic relaxation modelled by tunneling to a metallic reservoir, with $\gamma$ being the tunneling rate and metallic Green's function given by
\begin{equation}
		Q_\text{res}(\eps) = 
		\begin{pmatrix}
			1 &  2F(\eps)\\ 
			0 & 1
		\end{pmatrix}_\text{K} 
		\otimes \tau_3 ,
\end{equation}
where $F(\eps)=\tanh(\eps/2T) = 1-2n_F(\eps)$, with $n_F(\eps)$ being the Fermi distribution function. Such a model is equivalent to the relaxation-time approximation used in Refs.\ \cite{Eliashberg1970, Spivak1}.

In general, the order parameter field is characterized by two complex components: classical $\Delta(t)$ and quantum $\Delta_\text{q}(t)$. 
Because a homogeneous electric field couples exclusively to the gauge-invariant total momentum, it does not excite phase fluctuations (plasmons \cite{MooijShoen,Karuzin} and the neutral Bogoliubov-Anderson mode \cite{CG,Millis}). Hence, the fields $\Delta(t)$ and $\Delta_\text{q}(t)$ can be chosen real. This choice of the gauge is reflected in Eqs.\ \eqref{eqn:3:usadel} and \eqref{Sigma-def}, with the latter containing only one Pauli matrix $\tau_1$ multiplied by
\be
\label{check-Delta}
  \check{\Delta} = \Delta \sigma_0 + \Delta_\text{q} \sigma_1.
\ee

The real field $\ba(t)$ also comprises two components: classical $\ba(t)=e\mathbf{A}(t)/\hbar c$, determined by the electromagnetic vector potential \eqref{A(t)}, and its quantum counterpart $\ba_\text{q}(t)$, acting as a source for calculating electric current [see Eq.\ \eqref{j-def}]. Both components are combined into a matrix $\check\ba$ in the Keldysh space, in analogy with Eq.~\eqref{check-Delta}.

\subsection{Stationary supercurrent state}

In the stationary case with a time-independent $\mathbf{a}$, the saddle-point solution for the $Q$ matrix is diagonal in the energy space: 
\be
\label{Q-diag-stat}
  Q_{\eps_1,\eps_2} 
  = 2\pi \delta(\eps_1-\eps_2) Q(\eps_2) ,
\ee
with the standard triangular structure in the Keldysh space \cite{KA99,FLS2000}:
\be
  Q(\eps) = \begin{pmatrix}
			Q^\text{R}(\eps) & Q^\text{K}(\eps) \\ 
			0 & Q^\text{A}(\eps)
		\end{pmatrix}_\text{K} ,
\label{eqn:3:QMatrixKeldyshGeneral}
\ee
where $Q^\text{K}(\eps) = F(\eps) [Q^\text{R}(\eps)-Q^\text{A}(\eps)]$.
In the gauge with the real order parameter and the superfluid momentum determined by $\mathbf{a}$, the retarded (advanced) block can be written as
\begin{equation}
        Q^\text{R,A}(\eps) 
        = 
        \begin{pmatrix}
            g^\text{R,A}_\eps & f^\text{R,A}_\eps
            \\
            f^\text{R,A}_\eps & -g^\text{R,A}_\eps
        \end{pmatrix}_\text{N} ,
\end{equation}
where the normal ($g$) and Gor'kov ($f$) Green functions obey the symmetry relations
following from $Q^\text{A}=-\tau_3 Q^{\mathrm{R}\dagger}\tau_3$:
\be
\label{g-R-A}
  g^\text{A}_\eps = - (g^\text{R}_\eps)^* ,
\qquad
  f^\text{A}_\eps = (f^\text{R}_\eps)^* .
\ee
They can be conveniently represented in terms of a complex spectral angle $\theta^\text{R,A}_\eps$ as
\be
\label{g-f-theta}
  g^\text{R,A}_\eps = \pm \cos\theta^\text{R,A}_\eps ,
\qquad
  f^\text{R,A}_\eps = \pm \sin\theta^\text{R,A}_\eps ,
\ee
with the symmetry 
$\theta^\text{R}_{\eps} = -\theta^\text{A}_{-\eps} = -(\theta^\text{A}_\eps)^*$.

The spectral angle should be determined from the saddle-point (Usadel) equation $[Q,\delta S/\delta Q]=0$, yielding 
\begin{equation}
\label{Usadel-theta}
    i \eps^\text{R} \sin\theta^\text{R}_\eps + \Delta \cos\theta^\text{R}_\eps - \Gamma \sin\theta^\text{R}_\eps \cos\theta^\text{R}_\eps = 0,
\end{equation}
where $\eps^\text{R,A} = \eps \pm i \gamma/2$, $\Delta$ is the value of the order parameter, and $\Gamma$, introduced in Eq.~\eqref{eqn:2:GammaDef}, is the pair-breaking rate due to the supercurrent \cite{Maki}.
At zero current ($\Gamma=0$), 
$\theta^\text{R}_\eps = \arctan(i\Delta/\eps^\text{R})$ that corresponds to the BCS Green functions
\be
\label{gf-BCS}
  \begin{pmatrix} 
    g^\text{R,A}_\eps \\ f^\text{R,A}_\eps
  \end{pmatrix}
  = 
  \frac{\pm1}{\sqrt{(\eps\pm i\gamma/2)^2-\Delta^2}}
  \begin{pmatrix} 
    \eps\pm i\gamma/2 \\ i\Delta
  \end{pmatrix} ,
\ee
with the natural choice of the square-root branch at $\eps\to+\infty$.

Equation \eqref{Usadel-theta} incorporates two distinct mechanisms that smear the BCS density of states, controlled by the rates $\Gamma$ and $\gamma$. A finite $\Gamma$ (at $\gamma=0$) maintains a hard gap, while renormalizing its value from $\Delta$ to $E_g = (\Delta^{2/3}-\Gamma^{2/3})^{3/2}$ and cutting the BCS singularity
\cite{AG1960, AnthorePotier2003}. A finite $\gamma$ (at $\Gamma=0$) acts as the Dynes parameter smearing a hard gap \cite{Dynes}. When both rates are finite, the density of states is nonzero everywhere, 
but the peak shape is governed by the relation between $\Gamma$ and $\gamma$.

The order parameter $\Delta$ satisfies the self-consistency equation
\begin{equation}
\label{SCE}
\Delta
= 
\frac{\lambda}{2}\int d \eps \, 
F(\eps) \im \sin\theta^\text{R}_\eps ,
\end{equation}
which should be solved simultaneously with the Usadel equation \eqref{Usadel-theta}.

\subsection{Linear response}
\label{SS:linresp}

In the presence of a monochromatic microwave drive described by Eq.~\eqref{A(t)},
the order parameter becomes a periodic function that can be expanded in the Fourier series
\be \label{eqn:3c:DeltaExpansion}
  \Delta(t) = \sum_n \Delta^{(n)} e^{-in\omega t} ,
\ee
and the quasiparticle matter field acquires a form
\be
\label{Q-lattice}
  Q_{\eps_1,\eps_2} 
  = 2\pi \sum_{n} \delta(\eps_1-\eps_2-n\omega) 
  {\cal Q}^{(n)}_{\eps_1,\eps_2} .
\ee
In the absence of irradiation, only ${\cal Q}^{(0)}_{\eps,\eps}$ is nonzero, coinciding with $Q(\eps)$ introduced in Eq.\ \eqref{Q-diag-stat}.
The $n$th term in Eq.\ \eqref{Q-lattice} is responsible for quasiparticle transitions with the absorption of $n$ photon quanta and appears in the $n$th order of the perturbation theory in the ac field amplitude $\mathbf{A}_1$ \cite{Semenov2016}.
The calculation of a nonlinear response beyond perturbation theory is greatly complicated by the constraint $Q^2 = 1$, implying that different $\mathcal{Q}^{(n)}$ are not independent.

The current flowing through the system is calculated as a derivative with respect to the quantum component of the vector potential \cite{KA99,FLS2000}:
\be
\label{j-def}
  \mathbf{j}(\Omega) 
  = 
  -\frac{ e}{2V} 
  \left. \frac{\delta S}{\delta \mathbf{a}_q(-\Omega)} \right|_{a_q=0} .
\ee
For a periodic drive \eqref{A(t)}, $\Omega$ can be a multiple of the pumping frequency $\omega$. 

Hereafter we will be interested in the current at frequency $\omega$ linear in the microwave amplitude $\mathbf a_1$.
Using Eqs.\ \eqref{eqn:3:usadel} and \eqref{j-def} with $Q$ given by Eq.\ \eqref{Q-lattice}, we find two contributions to the current:
\be
\label{j+j}
  \mathbf{j}(\omega) = \mathbf{j}_1(\omega) + \mathbf{j}_2(\omega).
\ee
where
\begin{subequations}
\label{eqn:2:current}
\be
\label{eqn:2:drudeCurrent}
  \mathbf{j}_1(\omega)
  =
  \frac{\pi i \nu e D}{2} 
  { \mathbf a_1(\omega)} 
  \int \tr\left(\tau_3 Q_{\eps+\omega} \sigma_1 \tau_3 Q_{\eps}
  \right)
  (d\eps) 
\ee
and
\begin{multline}
  \mathbf{j}_2(\omega)
  =
  \frac{\pi i \nu e D}{2} \mathbf a_0 \int 
  \bigl[
  \tr
  \tau_3 Q_{\varepsilon} \sigma_1 \tau_3 
  \mathcal Q^{(1)}_{\varepsilon+\omega, \varepsilon} 
\\{}
  + 
  \tr
  \tau_3 \mathcal Q^{(1)}_{\varepsilon,\varepsilon-\omega} 
  \sigma_1 \tau_3 Q_{\varepsilon}
  \bigr]
  (d\eps) .   
  \label{eqn:2:debyeCurrent}
\end{multline}
\end{subequations}

The first contribution $\mathbf j_1(\omega)$ originates when both $\mathbf a_q$ and $\mathbf a_1$ are taken from the same $\check{\mathbf{a}}^2$ term in the action~\eqref{eqn:3:usadel}. It is aligned with ${ \mathbf a_1(\omega)}$ and is expressed in terms of the \emph{static} Green's functions $Q_\eps$ given by Eq.\ \eqref{eqn:3:QMatrixKeldyshGeneral}. It corresponds to the first term $\sigma_1(\omega)$ in Eq.\ \eqref{sigma} that generalizes the Drude conductivity to the superconducting case in the MB fashion \cite{MB,Fominov}.

The second contribution $\mathbf j_2(\omega)$ arises only in the presence of a dc supercurrent, i.e., finite $\mathbf a_0$, which multiplies $\mathbf a_q$ in the $\check{\mathbf{a}}^2$ term in the action. This contribution corresponding to $\sigma_2(\omega)$ in Eq.\ \eqref{sigma} is essentially nonequilibrium and contains a dynamic response of quasiparticles $\mathcal Q^{(1)}_{\varepsilon+\omega, \varepsilon}$ to the ac perturbation. In the linear order considered it suffices to write
\be
\label{dQ1/da1}
  \mathcal Q^{(1)}_{\varepsilon+\omega, \varepsilon}
  =
  \left(
    \frac{\partial \mathcal Q^{(1)}_{\varepsilon+\omega, \varepsilon}}{\partial{ \mathbf a_1(\omega)}}
    +
    \frac{\partial \mathcal Q^{(1)}_{\varepsilon+\omega, \varepsilon}}{\partial \Delta^{(1)}}
    \frac{\partial \Delta^{(1)}}{\partial{ \mathbf a_1(\omega)}}
  \right)
  { \mathbf a_1(\omega)} ,
\ee
where the derivatives should be taken at ${ \mathbf a_1(\omega)} = 0$.
Their explicit form can be found in Appendix \ref{app:CorrectionsToQ}.

The first term in Eq.\ \eqref{dQ1/da1} accounts for the microwave-induced modification of quasiparticle Green's functions. It involves dynamic diffusons $\mathcal{D}_{\eps+\omega,\eps}$, governing nonequilibrium corrections to the distribution function $F$, and cooperons $\mathcal{C}_{\eps+\omega,\eps}$, responsible for the change of $G^\text{R,A}$.

The second term in Eq.\ \eqref{dQ1/da1} captures the dynamics of the order parameter and its consequent feedback on quasiparticles. While the latter is again expressed in terms of diffusons and cooperons, the response of $\Delta(\omega)$ to ${ \mathbf a_1(\omega)}$ is governed by the retarded fluctuation propagator $L^\text{R}(\omega)$, now more commonly referred to as the Schmid-Higgs susceptibility $\chi_\text{SH}(\omega)$ \cite{Moor2017, Burmi2025, DzKamenev2025}:
\be
\label{eqn:2:delta1}
  \Delta^{(1)} 
  = 
  - (\pi i/4) D 
  L^\text{R}(\omega) \mathcal{J}(\omega) 
  \mathbf a_0  { \mathbf a_1 } ,
\ee
with the explicit form of $\mathcal{J}(\omega)$ provided by Eq.\ \eqref{eqn:4:Jdef}.

While the term $\mathbf j_1(\omega)$ given by Eq.\ \eqref{eqn:2:drudeCurrent} is ready for immediate calculation, to find $\mathbf j_2(\omega)$ in Eq.\ \eqref{eqn:2:debyeCurrent} one has to determine $\mathcal Q^{(1)}_{\varepsilon+\omega, \varepsilon}$. The latter has been calculated in Refs.\ \cite{Ovchinnikov2, Moor2017, Li2024} within the perturbative analysis of the dynamic Usadel and self-consistency equations, and we will follow a different approach implementing the standard machinery of the sigma-model.

\section{General expression for $\sigma(\omega)$}
\label{S:eqs-gen}

With the calculation details relegated to Appendix~\ref{app:CorrectionsToQ}, here we present the general expression for the linear-response conductivity tensor defined through 
$j_\alpha(\omega) = i \omega \hbar \sigma_{\alpha \beta}(\omega)a_{1\beta}(\omega)/e$. Its tensor structure is captured by Eq.\ \eqref{sigma}, with $\sigma_1(\omega)$ and $\sigma_2(\omega)$ corresponding to $\mathbf j_1(\omega)$ and $\mathbf j_2(\omega)$ in Eq.\ \eqref{j+j}, respectively.

The MB-like conductivity $\sigma_1(\omega)$ is given by
\be
\label{sigma1-res}
  \frac{\sigma_1(\omega)}{\sigma_0}
  =
  1 + K^\text{RA}(\omega) + K^\text{R}(\omega) ,
\ee
with $\sigma_0 = 2 e^2 \nu D$ being the Drude conductivity in the normal state and (hereafter we use a short-hand notation $\eps_\pm=\eps\pm\omega/2$)
\begin{subequations}
\label{KRA&KR}
\begin{equation}
K^\text{RA}(\omega)
=
-\frac{1}{4\omega}\int d\eps \,
  (F_{\eps_+} - F_{\eps_-})
  (1 + g^\text{R}_{\eps_+}g^\text{A}_{\eps_-} -  f^\text{R}_{\eps_+}f^\text{A}_{\eps_-}),
\label{eqn:5:kra}
\end{equation}
\begin{equation}
K^\text{R}(\omega)
= \frac{1}{2\omega}\int d\eps \, F_{\eps_-} (
1- g^\text{R}_{\eps_+}g^\text{R}_{\eps_-} + f^\text{R}_{\eps_+}f^\text{R}_{\eps_-}
).
\end{equation}
\end{subequations}
In the normal state, $g^\text{R}_\eps=-g^\text{A}_\eps=1$ and $f^\text{R}_\eps=f^\text{A}_\eps=0$, so that both $K^\text{RA}(\omega)$ and $K^\text{R}(\omega)$ vanish identically and we recover the Drude conductivity.
In the superconducting state, our result for $\sigma_1(\omega)$ given by Eqs.\ \eqref{sigma1-res} and \eqref{KRA&KR} coincides with the optical conductivity in the presence of paramagnetic impurities, see Eqs.\ (47) and (B14) of Ref.\ \cite{Fominov} [one should use Eq.\ \eqref{g-R-A} to bring Eqs.\ \eqref{KRA&KR} to the latter form].

According to Eq.\ \eqref{sigma2-2}, the contribution $\sigma_2(\omega)$ arising in the presence of the condensate superflow is a sum of $\sigma_2^\text{qp}(\omega)$ and $\sigma_2^\text{SH}(\omega)$, which correspond to the two terms in Eq.\ \eqref{dQ1/da1}, respectively.
The quasiparticle part of $\sigma_2(\omega)$ is given by
\be
\label{eqn:3:sigma2Generalized}
\frac{\sigma_2^\text{qp}(\omega)}{\sigma_0}
  = 
  - 
  \frac{\Gamma}{2T} 
  [\mathcal I_d(\omega) + 2 \mathcal I_c(\omega)] ,
\ee
where $\mathcal{I}_{d(c)}$ comes from the diffuson (cooperon) channel:
\begin{subequations}
\begin{multline}
\label{eqn:3:Idef}
  \mathcal{I}_d(\omega)
  = 
  \frac{ T}{\omega} \int d \varepsilon \, 
  (F_{\varepsilon_+}-F_{\varepsilon_-})
  \, \mathcal{D}_{\varepsilon_+,\varepsilon_-} 
\\{}
  \times 
  \bigl( g^\text{R}_{\varepsilon_+}f^\text{A}_{\varepsilon_-}
  + f^\text{R}_{\varepsilon_+}g^\text{A}_{\varepsilon_-}
  \bigr)^2
  \bigl(
  1 + g^\text{R}_{\varepsilon_+}g^\text{A}_{\varepsilon_-} + f^\text{R}_{\varepsilon_+}f^\text{A}_{\varepsilon_-}
  \bigr) ,
\end{multline}
\vspace{-10pt}
\begin{multline}
    \mathcal{I}_c(\omega) = \frac{T}{\omega} 
    \int d \varepsilon \, F_{\varepsilon_-} \, \mathcal{C}^\text{R}_{\varepsilon_+,\varepsilon_-} 
\\{} 
   \times
   \bigl( g^\text{R}_{\varepsilon_+}f^\text{R}_{\varepsilon_-}  
     + f^\text{R}_{\varepsilon_+}g^\text{R}_{\varepsilon_-}
   \bigr)^2
  \bigl(
  1 + g^\text{R}_{\varepsilon_+}g^\text{R}_{\varepsilon_-} + f^\text{R}_{\varepsilon_+}f^\text{R}_{\varepsilon_-}
  \bigr) .
    \label{eqn:4:ic}
\end{multline}
\end{subequations}
The contribution to $\sigma_2(\omega)$ due to the SH  mode excitation has the form
\begin{equation}
\label{sigma2d-gen}
   \frac{\sigma_2^\text{SH}(\omega)}{\sigma_0}
   =
   -i
   \frac{\pi^2 \Gamma}{32 \omega}
	 L^\text{R}(\omega) \mathcal{J}^2(\omega),
\end{equation}
where 
\begin{equation}
    \mathcal{J}(\omega) 
    = \mathcal{J}_{d}(\omega) + 2\mathcal{J}_{c}(\omega)
	  , 
\label{eqn:4:Jdef}
\end{equation}
with the diffuson and cooperon contributions given by
\begin{subequations}
\begin{multline}
  \mathcal{J}_d(\omega) 
  =
  4\int \frac{d \varepsilon}{2\pi} \,
  ( F_{\varepsilon_+} - F_{\varepsilon_-} )
  \,
  \mathcal{D}_{\varepsilon_+,\varepsilon_-}
\\{}
  \times
  \bigl(
    f^\text{R}_{\varepsilon_+} + f^\text{A}_{\varepsilon_-}
  \bigr) 
  \bigl(
  1 + g^\text{R}_{\varepsilon_+} g^\text{A}_{\varepsilon_-} - f^\text{R}_{\varepsilon_+} f^\text{A}_{\varepsilon_-}
  \bigr)
  ,
            \label{eqn:3:jdDef}
\end{multline}
\vspace{-10pt}
\begin{multline}
  \mathcal{J}_c(\omega) 
  = 
  4\int \frac{d \varepsilon }{2\pi} \,
  F_{\varepsilon_-} \,
  \mathcal{C}^\text{R}_{\varepsilon_+,\varepsilon_-}
\\{}
\times  
  \bigl(
f^\text{R}_{\varepsilon_+} + f^\text{R}_{\varepsilon_-}
  \bigr) 
  \bigl(
  1 + g^\text{R}_{\varepsilon_+} g^\text{R}_{\varepsilon_-} - f^\text{R}_{\varepsilon_+} f^\text{R}_{\varepsilon_-}
  \bigr)
  .
\end{multline}
    \label{eqn:3:jDef}
\end{subequations}

The formulae above involve the dynamic diffuson $\mathcal{D}_{\eps\eps'}$ and cooperon $\mathcal{C}^\text{R}_{\eps\eps'}$
defined as
\begin{subequations}
\label{eqn:2:diffCoop}
\begin{gather}
\label{D-def}
  \mathcal{D}_{\eps\eps'}
  =
  \frac{1}{\mathcal{E}^\text{RA}_{\eps\eps'} 
+ \Gamma 
  \bigl(
  1 + g^\text{R}_{\eps}g^\text{A}_{\eps'} + f^\text{R}_{\eps}f^\text{A}_{\eps'}
  \bigr)
  \bigl(
    g^\text{R}_{\eps}g^\text{A}_{\eps'} - f^\text{R}_{\eps}f^\text{A}_{\eps'}
  \bigr)
  }
  ,
\\
  \mathcal{C}^\text{R}_{\eps\eps'}
  =
  \frac{1}{\mathcal{E}^{RR}_{\eps\eps'}
+ \Gamma 
  \bigl(
  1 + g^\text{R}_{\eps}g^\text{R}_{\eps'} + f^\text{R}_{\eps}f^\text{R}_{\eps'}
  \bigr)
  \bigl(
    g^\text{R}_{\eps}g^\text{R}_{\eps'} - f^\text{R}_{\eps}f^\text{R}_{\eps'}
  \bigr)
  }
  ,
\end{gather}
\end{subequations}%
where $\mathcal{E}^{\alpha\alpha'}_{\eps\eps'} = \mathcal{E}^\alpha_\eps + \mathcal{E}^{\alpha'}_{\eps'}$ with $\alpha=\text{R}, \text{A}$ and 
\begin{equation}
    \mathcal{E}^\alpha_{\eps} 
    = \Delta f^\alpha_\eps - i(\eps \pm i \gamma/2 ) g^\alpha_\eps .
\end{equation}

The retarded fluctuation propagator, which determines $\sigma_2^\text{SH}(\omega)$ according to Eq.\ \eqref{sigma2d-gen}, has the form
\be
  L^\text{R}(\omega)
  =
  \left[
  -1/\lambda
  + \Pi_d(\omega) + \Pi_c(\omega) 
  \right]^{-1},
\label{eqn:2:propSum}
\ee
with the two contributions to the Cooper-channel susceptibility:
\begin{subequations}
\begin{multline}
\label{Pid-def}
  \Pi_d(\omega) =  -\frac{i}{4} \int d\varepsilon \, (F_{\varepsilon_+} - F_{\varepsilon_-} ) 
  \, \mathcal{D}_{\varepsilon_+,\varepsilon_-}
  \\
  \times 
  \bigl(
  1 + g^\text{R}_{\varepsilon_+} g^\text{A}_{\varepsilon_-} - f^\text{R}_{\varepsilon_+} f^\text{A}_{\varepsilon_-}
  \bigr)
   ,
   \end{multline}
\vspace{-10pt}
\begin{gather}
\label{Pic-def}
  \Pi_c(\omega) = -  \frac{i}{2} \int d\varepsilon \, F_{\varepsilon_-} 
  \, \mathcal{C}^\text{R}_{\varepsilon_+,\varepsilon_-}
\bigl(
  1 + g^\text{R}_{\varepsilon_+} g^\text{R}_{\varepsilon_-} - f^\text{R}_{\varepsilon_+} f^\text{R}_{\varepsilon_-}
  \bigr) 
   .
\end{gather}
\label{eqn:2:FluctPropGen}
\end{subequations}

\section{Conductivity near $T_c$}
\label{S:Tc}

In this Section, we analyze the behavior of $\sigma_{1,2}(\omega)$ in the Ginzburg-Landau limit, $T\to T_c$, where the general expressions of Sec.\ \ref{S:eqs-gen} can be evaluated analytically.

\emph{In the absence of a supercurrent}, the order parameter is given by the Ginzburg-Landau expression
\be
\label{DeltaGL}
  \Delta_0(T)
  =
  \sqrt{8 \pi^2/7 \zeta(3)} \sqrt{T_c (T_c - T)}
\ee
and the MB conductivity, $\sigma_\text{MB}(\omega)$, takes the form of Eq.~\eqref{sigma1-res} with
\begin{subequations}
\label{KMB}
\begin{gather}
\label{KMBRA}
  K^\text{RA}_{\Gamma=0}(\omega)
  = 
  \frac{\Delta}{2T} 
  \left[
    K(k') - E(k') + i E(k)
  \right] ,
\\{}
\label{KR-GL}
  K^\text{R}_{\Gamma=0}(\omega)
  = 
  \frac{i\pi\Delta^2}{2T\omega} .
\end{gather}
\end{subequations}
Here $K$ and $E$ are complete elliptic integrals of the first and second kind, respectively, with the modulus $k=\omega/2\Delta$ and complementary modulus $k'=\sqrt{1-k^2}$.
Hereafter analytic continuation from $\omega<2E_g$ to $\omega>2E_g$ is made assuming $\im\omega>0$.
The dissipative part, $\re\sigma(\omega)$, is almost equal to the Drude conductivity, with a small correction proportional to $\Delta_0/T$, which grows logarithmically at small $\omega$ due to the overlap of two BCS singularities.
The imaginary part,  $\im\sigma(\omega)$, demonstrates a usual superconducting response at low frequencies, in accordance with Eq.\ \eqref{Im-sigma-SC}.
Note that Eqs.\ \eqref{DeltaGL} and \eqref{KMB} are written in the leading order in $\Delta_0/T$.

\emph{In the presence of a supercurrent}, the order parameter is suppressed \cite{Tinkham}:
\be \label{eqn:5:DeltaGammaGL}
  \Delta^2 = \Delta_0^2 (1-\Gamma/3\Gamma_c) ,
\ee
where the critical depairing rate
\be
\label{eqn:5:GammaC}
  \Gamma_c 
  = 
  (4/3 \pi) (T_c-T) \ll \Delta
\ee
is much smaller than the order parameter for all currents [$I\propto\sqrt\Gamma\Delta^2$, see Eq.~\eqref{eqn:appE:currGen}] up to the critical current. 
As a result, the minigap $E_g=(\Delta^{2/3}-\Gamma^{2/3})^{3/2}$ is parametrically close to $\Delta$, with the coherence peak smearing
\be
  \omega_* \sim \Delta-E_g \sim \Gamma^{2/3}\Delta^{1/3} \ll \Delta
\ee
marking a crossover frequency that separates different $\omega$ regimes. 
The other relevant scale is set by the single-particle inelastic relaxation rate $\gamma$, which is assumed to be smaller than $\Gamma$. The crossover between the quasistatic and dynamic regimes in $\sigma_2^\text{qp}(\omega)$ takes place at $\omega\sim\gamma$. 
The crossover in $\sigma_2^\text{SH}(\omega)$ is determined by a much smaller rate $\gamma_Q\sim[\Delta(T)/T]\, \gamma$, see Eq.\ \eqref{gammaQ-def}.

The hierarchy of energy scales outlined in Eq.\ \eqref{eqn:5:GammaC} enables the analytical calculation of conductivity for arbitrary currents $I<I_c$ near $T_c$. Since $\Gamma\ll\Delta$, the Green functions are close to their bare BCS values \eqref{gf-BCS} everywhere except for the vicinity of the minigap, $\eps-E_g\sim\omega_*$.
According to the general scheme outlined in Sec.\ \ref{S:eqs-gen}, the conductivity contributions $\sigma_{1,2}(\omega)$ are expressed in terms of
six blocks, which are naturally grouped into two families: 
RA ($K^\text{RA}$, $\mathcal I_d$, $\mathcal J_d$)
and RR ($K^\text{R}$, $\mathcal I_c$, $\mathcal J_c$), each given by an energy integral.
Due to different analytical structures of the corresponding integrands, both families are expanded in $\Gamma$ in different ways.

For the RR family, the integration contour over $\eps$ can be deformed to reach Matsubara poles in the upper half-plane at energies $2i\pi T(n+1/2)$. Therefore the relative correction to the BCS ($\Gamma=0$) values of $K^\text{R}$, $\mathcal I_c$, $\mathcal J_c$ will be of the order of $\Gamma/T$, which is parametrically smaller than corrections arising in the RA-family blocks [see Eq.~\eqref{eqn:5:KRAinterm}, where $\delta K^\text{RA}/K^\text{RA}_{\Gamma=0} \sim \Gamma/\omega$].
Evaluating RR blocks at zero current, we obtain
\be
\label{Ic-Jc-GL}
  \mathcal{I}_c(\omega)
  = 
  \frac{14 i  \zeta(3)}{\pi^2} \frac{\Delta^2}{T\omega} ,
\qquad
  \mathcal{J}_c(\omega)
  =
  \frac{2i \Delta}{T} .
\ee
Equations \eqref{KR-GL} and \eqref{Ic-Jc-GL} for RR blocks will be used below to calculate conductivity in various frequency domains in the vicinity of $T_c$.

The behavior of the RA-family blocks at small $\Gamma$ depends on the relation between $\omega$ and $\omega_*$, and will be analyzed below.

\subsection{Separated peaks regime, $\omega\gg\omega_*$}

In this frequency range, two smeared BCS peaks located at $\varepsilon \approx \pm \Delta \pm \omega/2$ do not overlap. Therefore the $\eps$-integration contour can be moved away from the branching points $\pm E_g\pm\omega/2$ to the region, where the Green functions almost coincide with their bare BCS values~\eqref{gf-BCS}. Thus we conclude that in this regime the RA-family blocks can be calculated using a formal perturbation theory in the parameter $\Gamma/\Delta$ on top of the zero-current state. In particular, the first-order correction to the bare Green functions following from Eq.\ \eqref{Usadel-theta} has the form
\be
\label{dg-df}
  \delta g^\text{R}_\eps = (\Gamma/\Delta) (f^\text{R}_\eps)^3 g^\text{R}_\eps ,
\qquad
  \delta f^\text{R}_\eps = -(\Gamma/\Delta) (f^\text{R}_\eps)^2 (g^\text{R}_\eps)^2 ,
\ee
and analogously for $g^\text{A}_\eps$ and $f^\text{A}_\eps$.
The absence of nonanalytic terms in $\Gamma$ (the independence of the integral on a particular peak smearing mechanism) follows from the possibility of the contour deformation discussed above.

To get the result in the leading order in $\Gamma$, we write 
$K^\text{RA}(\omega)=K^\text{RA}_{\Gamma=0}(\omega)+\delta K^\text{RA}(\omega)$, where the correction is obtained by substituting Eq.\ \eqref{dg-df} into Eq.\ \eqref{eqn:5:kra}:
\be
\label{eqn:5:KRAinterm}
  \delta K^\text{RA}(\omega) = 
    \frac{\pi \Gamma \Delta}{8 T \omega}  \left( \frac{\omega-\Delta}{\sqrt{\omega(2\Delta+\omega)}} - i \frac{\omega+\Delta}{\sqrt{\omega(2\Delta-\omega)}}
    \right) .
\ee
Since Eqs. \eqref{eqn:3:sigma2Generalized} and \eqref{sigma2d-gen} for $\sigma^\text{qp}_2(\omega)$ and $\sigma^\text{SH}_2(\omega)$ already contain $\Gamma$, the corresponding RA blocks $\mathcal I_d$ and $\mathcal J_d$ can be calculated with the BCS Green functions \eqref{gf-BCS} that yields
\begin{gather}
\label{eqn:5:IInterm}
  \mathcal{I}_d(\omega)
  =
  \pi \frac{\Delta}{\omega}
    \left(
      \frac{\Delta+{\omega}}{\sqrt{\omega(2\Delta+{\omega})}} 
      - i \frac{\Delta-{\omega}}{\sqrt{{\omega}(2\Delta-{\omega})}}
    \right) ,
\\
  \mathcal{J}_d(\omega)
  =
  -\frac{2i\Delta}{T} 
  \left(
    1
    + \frac{i\omega}{\pi\Delta}
	[ K(k')+ i K(k) ]
  \right) . \label{eqn:5a1:jd}
\end{gather}

The results obtained above together with the fluctuation propagator $L^\text{R}$ given by \eqref{eqn:2:fluctPropTTC} make it possible to calculate three conductivity contributions according to Eqs.\ \eqref{sigma1-res}, \eqref{eqn:3:sigma2Generalized} and \eqref{sigma2d-gen} in the first order in $\Gamma$. We present them in the most interesting limit of subgap frequencies, $\omega\ll\Delta$:
\begin{subequations}
\begin{gather}
\label{sigma1-large-w}
  \frac{\sigma_1(\omega)}{\sigma_0} 
  = 
  1 + \frac{\Delta}{2T} \log \frac{8\Delta}{e\omega} 
  + \frac{i\pi \Delta^2}{2 T \omega} 
  - \frac{e^{i\pi/4}\pi \Gamma \Delta^{3/2}}{8 T \omega^{3/2} },
\\
  \frac{\sigma_2^{\mathrm{qp}}(\omega)}{\sigma_0} 
  =
  - \frac{e^{-i\pi/4}\pi \Gamma \Delta^{3/2}}{2 T \omega^{3/2}},
\\
  \frac{\sigma_2^{\mathrm{SH}}(\omega)}{\sigma_0} 
  = 
  - \frac{\Gamma}{T} \left( \log\frac{8\Delta}{\omega} 
  + \frac{i\pi \Delta}{2\omega}
    \right).
\end{gather}
\end{subequations}
Our Eq.\ \eqref{sigma1-large-w} for $\Gamma=0$ (zero current) coincides with the low-frequency expansion of the MB conductivity, as given by Eq.~(47) of Ref.~\cite{Ovchinnikov1}. The terms proportional to $\Gamma$ provide a correction to the MB result, with $\re\delta\sigma(\omega)/\sigma_0 \sim (\Delta/T)(\omega_*/\omega)^{3/2}$, coming from both $\sigma_1(\omega)$ and $\sigma_2^\text{qp}(\omega)$. This correction grows with decreasing frequency and becomes comparable to $\re\sigma_\text{MB}(\omega)/\sigma_0-1$ at $\omega\sim\omega_*$, right at the border of applicability of the separated peaks regime. Note that in this regime the contribution  of $\sigma_2^{\mathrm{SH}}$ is small and can be neglected.

\subsection{Overlapping peaks regime, $\omega\ll\omega_*$}
\label{SSS:overlapping}

\subsubsection{Derivation}

In this frequency range, the smeared coherence peaks at $\Delta\pm\omega/2$ become practically indistinguishable. That leads to modification of the expressions for the RA-family blocks $K^\text{RA}$, $\mathcal{I}_d$, and $\mathcal{J}_d$ obtained above, while expressions for the RR-family blocks given by Eqs. \eqref{KR-GL} and \eqref{Ic-Jc-GL} remain unchanged.

With logarithmic accuracy, the block $K^\text{RA}$ is given by
\be
  K^\text{RA}(\omega) 
  = 
  \frac{ \Delta}{3 T} \log \frac{\Delta}{\Gamma} ,
\ee
that can be understood as the usual MB log singularity $K^\text{RA}(\omega) = (\Delta/2T) \ln(\Delta/\omega)$ [see Eqs.\ \eqref{KMB} and \eqref{sigma1-large-w}] cut at the peak width $\omega\sim\omega_*$.

Calculating RA blocks $\mathcal{I}_d$ and $\mathcal{J}_d$ is more challenging, as they contain the diffuson propagator $\mathcal{D}_{\varepsilon_+,\varepsilon_-}$ with nearly equal energies.
In the absence of the supercurrent ($\Gamma=0$), the diffuson denominator in Eq.\ \eqref{D-def} at coincident energies and at $\gamma=0$ is given by $\mathcal{D}_{\eps,\eps}^{-1}=2\re\sqrt{\Delta^2-\eps^2}$, exactly vanishing above the gap. The same remains true for $\Gamma>0$ as well:  $\mathcal{D}_{\eps,\eps}^{-1}=0$ for $\eps>E_g$. A finite $\omega + i\gamma$ regularizes the otherwise divergent diffuson propagator above the gap, leading to the expression
\be
\label{dif-M}
  \mathcal{D}_{\eps_+,\eps_-}
  \approx
  \frac{\Theta(|\eps|-E_g)}{-i M(\eps) (\omega+i\gamma)} ,
\ee
where the function $M(\eps)$ is calculated in Appendix \ref{app:Asymp}.
In the considered regime ($\omega\ll\omega_*$), the energy integrations in Eqs.\ \eqref{eqn:3:Idef} and \eqref{eqn:3:jdDef} come from the peak region, $|\eps-E_g| \sim \omega_*$. We evaluate them in Appendix \ref{app:Asymp} and get 
\be
\label{Id-overlap-res}
  \mathcal{I}_d(\omega)
  = 
  -2i 
  \beta_\text{qp} 
  \frac{\Delta^{1/3}}{\Gamma^{1/3}}
  \frac{\Delta}{\omega+ i \gamma} ,
\ee
where 
\be
\label{beta-qp}
  \beta_\text{qp} 
  =
  \frac{3^{3/2} \pi^2}{5 \times 2^{4/3}\, \Gamma^3(2/3)} 
  = 1.639 .
\ee

The same procedure leads to a much simpler expression for the block $\mathcal{J}_d$:
\be
\label{Jd-overlap-res}
  \mathcal{J}_d(\omega)
  = 
  - \frac{2i\Delta}{T}
  \frac{\omega}{\omega+i\gamma} .
\ee
Note that this result can be formally obtained from the leading first term of Eq.~\eqref{eqn:5a1:jd} if 1 is replaced by $\omega/(\omega+i\gamma)$, with $\omega$ in the numerator coming from the term $F_{\eps_+}-F_{\eps_-}$ in Eq.\ \eqref{eqn:3:jdDef} and $\omega+i\gamma$ in the denominator coming from the pole of the diffuson \eqref{D-def}.

Collecting altogether and using the fluctuation propagator \eqref{LR-res-app}, we arrive at the following expressions for the conductivity contributions at $\omega\ll\omega_*$:
\begin{subequations}
\begin{gather}
\label{sigma1-w<w*}
    \frac{\sigma_1(\omega)}{\sigma_0 } = 1+ \frac{\Delta}{3T} \log\frac{\Delta}{\Gamma} + \frac{i \pi \Delta^2}{2 T \omega},
\\{}
  \frac{\sigma^\text{qp}_2(\omega)}{\sigma_0 }
  = 
  \beta_\text{qp} 
  \frac{ \Delta^{4/3} \Gamma^{2/3}}{(\gamma - i \omega ) T} ,
\\{}
 \frac{\sigma^\text{SH}_2(\omega)}{\sigma_0}
  = - 
  \frac{i\pi\Delta^2}{2 T \omega}
  \frac{\Gamma}{\Delta} \frac{\left( 2 - \frac{\omega}{\omega+i\gamma}  \right)^2}
  {\frac{\omega}{\omega+i\gamma} + \frac{2(3\Gamma_c-\Gamma)}{\Delta} }.
\label{eqn:5:sigmaSHAsympTTC}
\end{gather}%
\label{eqn:5:TTCcondTerms}%
\end{subequations}%
Our expressions for the conductivity in the overlapping peaks regime, Eqs.\ \eqref{eqn:5:TTCcondTerms}, precisely reproduce the earlier result of Ovchinnikov and Isaakyan \cite{Ovchinnikov2}, their Eq.\ (12).

\subsubsection{Analysis}

At the border of the overlapping peaks regime, $\omega\sim\omega_*$, $\sigma_1(\omega)$ is the leading contribution.
As the frequency decreases, the $\sigma_2(\omega)$ terms grow and, owing to a small $\gamma$ in the denominator, may start to compete with $\sigma_1(\omega)$.

Since the three conductivity contributions in Eqs.\ \eqref{eqn:5:TTCcondTerms} demonstrate different frequency dependence, $\sigma(\omega)$ in the $\bA_1{\parallel}\bA_0$ configuration, which is given by their sum, exhibits a cascade of crossovers with decreasing $\omega$. The corresponding crossover frequencies depend on the inelastic relaxation rate $\gamma$ and the supercurrent depairing rate $\Gamma$ in a complicated fashion.

The quasiparticle contribution $\re\sigma_2^\text{qp}(\omega)$ has a Lorentz shape with the width $\gamma$. 
The structure of the Schmid-Higgs term $\sigma^\text{SH}_2(\omega)$ is more intricate, with the characteristic frequency $\gamma$ in the numerator and $\gamma_Q\sim[\Delta(T)/T]\gamma$ [see Eq.\ \eqref{gammaQ-def}] in the denominator. The latter is much smaller than $\gamma$ in the limit $T\to T_c$ considered.
Hence it is the frequency scale $\gamma_Q$, which determines the apparent crossover between the quasistatic and dynamic regimes.

In the quasistatic regime ($\omega\ll\gamma_Q$), the dissipative conductivity saturates to
\be \label{eqn:5:ReSigma0TTc}
  \frac{\re\sigma(0)}{\sigma_0}
  =
  1
  +
  \beta_\text{qp} 
  \frac{ \Delta^{4/3} \Gamma^{2/3}}{\gamma T} 
  +
  \frac{\pi \Gamma \Delta^3}{2T\gamma (3\Gamma_c - \Gamma)^2} ,
\ee
with the three terms corresponding to $\sigma_1$, $\sigma_2^\text{qp}$ and $\sigma_2^\text{SH}$, respectively.
Since $\sigma_2(0)$ contain a small inelastic relaxation rate $\gamma$ in the denominator, it may exceed the bare Drude conductivity in a generic case. This topic has been recently brought up in Ref.\ \cite{Spivak1}, where the authors obtained a similar expression for $\sigma_2^\text{qp}(\omega)$, with their coefficient for $\beta_\text{qp}\approx0.087$ being 18 times smaller than our Eq.\ \eqref{beta-qp}. However, in Ref.\ \cite{Spivak1} the Schmid-Higgs contribution $\sigma_2^\text{SH}(\omega)$ was completely neglected.

The dissipative conductivity enhancement in the quasi-static limit is maximal at the critical current ($\Gamma=\Gamma_c$), where it is determined by the SH term, since the ratio $\re\sigma_s^\text{qp}(0)/\re\sigma_s^\text{SH}(0) \sim (\Gamma_c/\Delta)^{5/3} \sim (\Delta/T)^{5/3} \ll1$. Using Eqs.\ \eqref{DeltaGL} and \eqref{eqn:5:GammaC}, we obtain a huge enhancement of dissipation by inelastic relaxation at the critical current:
\be \label{eqn:5:maxReSigma0TTc}
  \left.\frac{\re\sigma(0)}{\sigma_0}\right|_{I=I_c}
  \sim
  \frac{\Delta}{\gamma} .
\ee
With decreasing $\omega$, $\re\sigma(\omega)$ reaches its maximal value of $\re\sigma(0)$ through a set of crossovers determined by the competition between the terms in Eqs.\ \eqref{eqn:5:TTCcondTerms}, as one can see in Fig.\ \ref{F:low-w-log}.

Finally, we discuss the behavior of $\im\sigma(\omega)$. It is determined by $\sigma_1(\omega)$ up to the lowest frequencies $\omega\sim\gamma_Q$, where the SH contribution becomes of the same order. In the quasistatic limit ($\omega\ll\gamma_Q$), the quantity $\Omega_s$ introduced in Eq.~\eqref{Im-sigma-SC} acquires the form
\be \label{eqn:5:OmegasTTc}
  \Omega_s
  =
  \frac{\pi\Delta^2}{2T} 
  \left( 1- \frac{2\Gamma}{3\Gamma_c - \Gamma} \right).
\ee
At the critical current ($\Gamma=\Gamma_c$), the SH contribution completely cancels the standard MB contribution, leading to the absence of the inductive response in the quasistatic limit. This analytic result obtained in the vicinity of $T_c$ is in full agreement with numerical observations presented in Fig.\ \ref{fig:coeffAAtZeroOmega}.

\section{Discussion}
\label{S:Discussion}

Having performed a comprehensive analysis based on the exact expression for $\sigma(\omega)$, we are now in a position to reconcile and critically reassess previously reported results for the optical conductivity of a current-carrying superconductor.

Conceptually, our theory aligns closely with the works of Ovchinnikov \cite{Ovchinnikov1} and Ovchinnikov and Isaakyan \cite{Ovchinnikov2}, who also performed a perturbative analysis of the quasiclassical equations for dirty superconductors. Their approaches involved Matsubara formalism followed by analytic continuation to real frequencies \cite{Ovchinnikov1} and Keldysh real-time technique \cite{Ovchinnikov2}. Using an alternative framework of the Keldysh sigma model, we arrive at exactly the same general expression for $\sigma(\omega)$. Moreover, we present a detailed derivation that clarifies the physical origin of each term. Beyond independently confirming the results Refs.\ \cite{Ovchinnikov1, Ovchinnikov2}, our analysis delineates the key features of the optical conductivity and traces their systematic evolution throughout the full $\omega$, $T$, and $I$ parameter space.

The giant low-frequency absorption due to inelastic relaxation addressed by Smith \emph{et al}.\ \cite{Spivak1} is a manifestation of the singular behavior $\sigma_2^\text{qp}(\omega)\propto1/(\omega+i\gamma)$ near $T_c$. However, the coefficient they reported in this expression is approximately 18 times smaller than the value calculated in Ref.\ \cite{Ovchinnikov2} and confirmed by us. The discrepancy can be attributed to the oversimplified kinetic equation employed in Ref.\ \cite{Spivak1}, which does not account for coherence factors. We also emphasize that for sufficiently large currents, the SH contribution $\re\sigma_2^\text{SH}(\omega)$, omitted in Ref.\ \cite{Spivak1}, dominates  $\re\sigma_2^\text{qp}(\omega)$ at lowest frequencies.

An unusual behavior of the optical conductivity at absolute zero has been recently reported by Kubo \cite{Kubo2025}, who obtained a sign change of $\Omega_s$ within the current interval $0.91<I/I_c(0)<0.99$. For moderate currents, $I/I_c(0)\lesssim0.8$, our results show excellent qualitative agreement with those of Ref.\ \cite{Kubo2025}, including the emergence of a frequency window with negative $\im\sigma(\omega)$. However, we show that $\Omega_s$, which is a measure of the superfluid density, stays positive for all currents up to $I_c(0)$, see inset to Fig.\ \ref{fig:coeffAAtZeroOmega}.
The function $\Omega_s(I)$ calculated in Appendix \ref{A:SF0} and obtained previously in Ref.\ \cite{Ovchinnikov1} vanishes precisely at the critical current, where the negative contributions of $\im\sigma_2^\text{qp}(\omega)$ and $\im\sigma_2^\text{SH}(\omega)$ exactly cancel the positive term $\im\sigma_1(\omega)$.
Note that the negative weight of the SH mode in the superfluid density has recently been discussed using sum-rule and gauge-invariance arguments~\cite{Wang2025}.

\section{Summary}
\label{S:Summary}

Using the Keldysh sigma model formalism, we have derived the general expression for an anisotropic optical conductivity $\sigma(\omega)$ of a dirty superconductor biased by a dc current. Our combined analytical and numerical study tracks the key features of the response across all frequencies, temperatures, and currents. At high frequencies, they include the emergence of a peak in $\re\sigma(\omega)$ and an associated dip in $\im\sigma(\omega)$ near $2E_g$, with the amplitude of both growing with increasing $I$ and decreasing $T$. Below $T\lesssim 0.75 \, T_c$ at large currents, the enhanced role of the Schmid-Higgs mode drives a sign change in the reactive conductivity, creating a frequency band with a capacitive rather than inductive response. At low frequencies, the main effects are the giant absorption governed by inelastic relaxation and sensitivity of the inductive response to the energy relaxation rate. Our analysis reveals crucial importance of both the quasiparticle relaxation channel and the amplitude Schmid-Higgs mode, whose fine tuning is responsible for the exact cancellation of the superfluid density at the critical current.

\acknowledgments

We thank I. S. Burmistrov for pointing out the role of the Schmid-Higgs mode.
This work was supported by the Russian Science Foundation under Grant No.\ 23-12-00297 and by the Foundation for the Advancement of Theoretical Physics and Mathematics ``BASIS''. 
A.P. acknowledges support from the Ministry of Science and Higher Education of Russian Federation (Project No.\ FFWR-2024-0017).

\appendix
\section{Sigma model \label{app:SigmaModel}}

\subsection{Diffusons and cooperons}

Fluctuations of the $Q$ matrix around the stationary superconducting saddle point \eqref{eqn:3:QMatrixKeldyshGeneral} can be conveniently parametrized as \cite{TSK18}
\begin{equation}
\label{Q-W}
  Q
  =
  \mathcal{U}^{-1} 
  \sigma_3 \tau_3 \left(1+ W + W^2/2 + \dots \right) 
  \mathcal{U},
\end{equation}
where $\mathcal{U}$ is given by 
\begin{equation}
    		\mathcal{U} = 
		\begin{pmatrix}
			e^{i \tau_2 \theta^\text{R}(\eps)/2} & 0 \\
			0 & e^{i \tau_2 \theta^\text{A}(\eps)/2}
		\end{pmatrix}_{\mathrm{K}} \otimes 
		\begin{pmatrix}
			1 &  F\\ 
			0 & -1
		\end{pmatrix}_{\mathrm{K}} 
\end{equation}
and $W$ is a matrix in $\text{K}\otimes\text{N}$ and energy spaces, which anticommutes with $\sigma_3 \tau_3$. In a general case, $W$ has $8$ nonzero components, see, e.g., Ref.\ \cite{Karuzin}. While a \emph{nonuniform} time-dependent vector potential $\mathbf{A}_1(\br,t)$ couples to all of them, only half of the diffusive modes are excited by a \emph{spatially uniform} microwave field.
Technically, this occurs because in the uniform case the full long-derivative part of the action, $\Tr(\nabla Q-i[\check{\mathbf{a}}\tau_3,Q])^2$, reduces to the zero-dimensional form of Eq.\ \eqref{eqn:3:usadel}. The latter lacks the mixed term $\Tr[\check{\mathbf{a}}\tau_3,Q]\nabla Q$, which is responsible for excitation of phase fluctuations and  charge imbalance. 
In this case, it is sufficient to use a simplified parametrization~\cite{TSK18}
\begin{equation}
    		W = \begin{pmatrix}
			c^\text{R} i \tau_2 & d \\
			\overline{d}  & c^\text{A} i \tau_2
		\end{pmatrix},
\end{equation}
where $c^\text{R}$ and $c^\text{A}$ are the cooperon modes, responsible for the modification of $\theta^\text{R}$ and $\theta^\text{A}$, accordingly, while $d$ and $\overline{d}$ are the diffuson modes, which describe perturbations of the distribution function $F$. 

At the Gaussian level, fluctuations of the $c$ and $d$ modes are characterized by the correlation functions
\begin{subequations}
\label{eqn:app1:corr}
\begin{gather}
  \corr{ c^{\alpha}_{\eps_1,\eps_2} c^{\alpha}_{\eps_3,\eps_4} }
  = 
  (\delta/\pi)
  \hat{\delta}(\eps_1-\eps_4) \hat{\delta}(\eps_2-\eps_3)
  \, \mathcal{C}^\alpha_{\eps_1,\eps_2},
\\
  \left \langle d_{\eps_1,\eps_2} \overline{d}_{\eps_3,\eps_4} \right \rangle  
  = 
  (\delta/\pi)
  \hat{\delta}(\eps_1-\eps_4) \hat{\delta}(\eps_2-\eps_3)
  \, \mathcal{D}_{\eps_1,\eps_2},
\end{gather}
\end{subequations}
where $\hat{\delta}(\eps-\eps') = 2\pi \delta(\eps-\eps')$, and the diffuson and cooperon propagator are given by Eqs.~\eqref{eqn:2:diffCoop}.

\subsection{Fluctuation propagator}

Denoting deviations of the order parameter from the mean-field value by $\Delta_1$, keeping the term in the action bilinear in $W$ and $\Delta_1$, evaluating the Gaussian integral over $W$, and combining the result with the last term in Eq.\ \eqref{eqn:3:usadel}, we arrive at the effective action
\be
  S^{(2)}[\mathbf{\Delta}_1] = \frac{2}{\delta} \int (d \omega) \mathbf{\Delta}_1(-\omega) L^{-1}(\omega) \mathbf{\Delta}_1(\omega),
\ee
where the vector $\mathbf{\Delta}_1 = (\Delta_1,\Delta_{1q})^T$ comprises the classical and quantum components. 
The fluctuation propagator, which is defined as the inverse of the matrix $L^{-1}$, has the standard bosonic matrix structure in the Keldysh space \cite{KA99,FLS2000}:
\be
  L(\omega) = \begin{pmatrix}
		L^\text{K}(\omega) & L^\text{R}(\omega) \\
		L^\text{A}(\omega) & 0
	\end{pmatrix} ,
\ee
with $L^\text{K}(\omega)=\coth(\omega/2T)[L^\text{R}(\omega)-L^\text{A}(\omega)]$.
It determines the correlation functions of the order parameter fluctuations:
\begin{subequations}
\begin{gather}
  \langle \Delta_1(\omega)\Delta_{1q}(-\omega)\rangle 
  = 
  (i \delta/4) L^\text{R}(\omega), 
\\
  \langle \Delta_1(\omega)\Delta_1(-\omega)\rangle 
  =
  (i \delta/4) L^\text{K}(\omega).
\end{gather}
\end{subequations}

The retarded component of the fluctuation propagator given by Eq. \eqref{eqn:2:propSum} is directly related to the dynamic susceptibility of the Schmid-Higgs mode$, L^\text{R}(\omega)=\chi_{\mathrm{SH}}(\omega)$, a subject of significant recent interest \cite{Burmi2025, DzKamenev2025}.

\subsection{Microwave-induced corrections to $W$}
\label{SS:W-corr}

Microwave irradiation modifies the stationary superconducting saddle point. As discussed in Sec.\ \ref{SS:linresp}, the linear in $\mathbf{A}_1$ corrections arise either directly, from the $\check{\mathbf{a}}^2$ term in the action~\eqref{eqn:3:usadel}, or indirectly, through modulation of the order parameter [see Eq.~\eqref{eqn:3c:DeltaExpansion}]:
\be
\label{W=dir+ind}
  \corr{W} = \corr{W}^\Gamma + \corr{W}^\Delta .
\ee
Since $\mathbf{A}_1$ is a classical field, both $\corr{W}$ and the induced order parameter $\Delta_1$ have no quantum components: $\corr{\overline d}=0$ and $\Delta^{(1)}_q=0$. 
The direct contributions are given by 
($\alpha=\text{R}$ or A)
\begin{subequations}
\be
  \corr{ c^{\alpha}_{\varepsilon,\varepsilon'} }^\Gamma
  = 
  - 4 D
  \sin\left( \theta^{\alpha}_\varepsilon + \theta^{\alpha}_{\varepsilon'}\right)
  \cos \frac{\theta^{\alpha}_\varepsilon - \theta^{\alpha}_{\varepsilon'}}{2} \, 
  \mathcal{C}^{\alpha}_{\varepsilon,\varepsilon'} 
  \, \mathbf{a}_0 \mathbf{a}_{1,\varepsilon-\varepsilon'} ,
\ee
\begin{multline}
  \corr{ d_{\varepsilon,\varepsilon'} }^\Gamma
  = 
  4 D
  \sin ( \theta^\text{R}_\varepsilon + \theta^\text{A}_{\varepsilon'} ) 
  \sin \frac{\theta^\text{R}_\varepsilon - \theta^\text{A}_{\varepsilon'}}{2} 
\\{}
  \times
  (F_{\varepsilon}-F_{\varepsilon'})
  \, \mathcal{D}_{\varepsilon,\varepsilon'} 
  \, \mathbf{a}_0 \mathbf{a}_{1,\varepsilon-\varepsilon'} .
\end{multline}
\end{subequations}
The indirect contributions are given by
\begin{subequations}
\label{eqn:app1:averages}
\begin{gather}
  \corr{ c^\text{R}_{\varepsilon,\varepsilon'} }^\Delta
  = 
  4 \cos\frac{\theta^\text{R}_{\eps} + \theta^\text{R}_{\eps'}}{2} 
  \, \mathcal{C}^\text{R}_{\eps,\eps'}
  \Delta_{1,\eps-\eps'} ,
\\
  \corr{ c^\text{A}_{\varepsilon,\varepsilon'} }^\Delta
  = 
  - 4 \cos\frac{\theta^\text{A}_{\eps} + \theta^\text{A}_{\eps'}}{2} 
  \, \mathcal{C}^\text{A}_{\eps,\eps'}
  \Delta_{1,\eps-\eps'} ,
\\
  \corr{ d_{\varepsilon,\varepsilon'} }^\Delta
  = -2 \sin \frac{\theta^\text{R}_{\eps} + \theta^\text{A}_{\eps'}}{2} 
    (F_{\eps}- F_{\eps'})
  \, \mathcal{D}_{\eps,\eps'}
  \Delta_{1,\eps-\eps'} ,
\end{gather}
\end{subequations}
where the first-order correction to the order parameter is determined by Eq.\ \eqref{eqn:2:delta1}.

Equations \eqref{W=dir+ind}--\eqref{eqn:app1:averages} provide the linear-in-$\mathbf{A}_1$ shift of the matrix $Q$ in the absence of loop (weak localization) corrections.

\section{Explicit expression for $\mathcal{Q}^{(1)}$}
\label{app:CorrectionsToQ}

The first-order nonstationary correction to the Green's function [Eq.\ \eqref{dQ1/da1}] can be obtained from the linear term in Eq.~\eqref{Q-W} as
\begin{equation}
    Q^{(1)}
    =
    \mathcal{U}^{-1} \sigma_3 \tau_3 \corr{W} \,\mathcal{U}
,
\end{equation}
where the microwave-induced correction $\corr{W}$ is calculated in Appendix \ref{SS:W-corr}. The first correction in Eq.\ \eqref{Q-lattice}, $\mathcal{Q}^{(1)}_{\varepsilon + \omega,\varepsilon}$, shares the triangular structure of Eq.~\eqref{eqn:3:QMatrixKeldyshGeneral}.
Its derivatives with respect to $\mathbf{a}_1$ are given by
($\alpha=\text{R}$ or A)%
\begin{subequations}
\label{eqn:app:dQ1/da1Explicit}
\be
  \frac{\partial \mathcal Q^{(1)\alpha}_{\varepsilon+\omega, \varepsilon}}{\partial{ \mathbf a_1(\omega)}}
  = 
  -
  {\mathbf a_0} D \, \mathcal{C}^\alpha_{\varepsilon+\omega, \varepsilon} m^\alpha_{\varepsilon+\omega, \varepsilon},
\ee
\begin{multline}
\label{eqn:app:dQ1/da1Explicit-K}
  \frac{\partial \mathcal Q^{(1)\text{K}}_{\varepsilon+\omega, \varepsilon}}{\partial  \mathbf a_1(\omega)}
  =
  \frac{\partial \mathcal Q^{(1)\text{R}}_{\varepsilon+\omega, \varepsilon}}{\partial  \mathbf a_1(\omega)} 
  F_{\varepsilon} 
  -
  F_{\varepsilon+\omega} 
  \frac{\partial \mathcal Q^{(1)\text{A}}_{\varepsilon+\omega, \varepsilon}}{\partial  \mathbf a_1(\omega)}
\\
  -
  { \mathbf a_0} D \, \mathcal{D}_{\varepsilon+\omega, \varepsilon} (F_{\varepsilon+\omega}-F_{\varepsilon})m^\text{K}_{\varepsilon+\omega, \varepsilon},
\end{multline}
\end{subequations}
where we denote
$m^\alpha_{\eps_1,\eps_2}
=
Q^\alpha_{\eps_1} T^{\alpha\alpha}_{\eps_1,\eps_2} Q^\alpha_{\eps_2}
- T^{\alpha\alpha}_{\eps_1,\eps_2}$,
$m^\text{K}_{\eps_1,\eps_2}
=
Q^\text{R}_{\eps_1} T^\text{RA}_{\eps_1,\eps_2} Q^\text{A}_{\eps_2} 
-
T^\text{RA}_{\eps_1,\eps_2} ,
$
and
$  T^{\alpha\beta}_{\eps_1,\eps_2}
  =
  \tau_3 
  (Q^\alpha_{\eps_1}+Q^\beta_{\eps_2}) \tau_3 
$.

The derivatives with respect to $\Delta^{(1)}$ are given by
\begin{subequations}
\label{eqn:app:dQ1/dDelta}
\be
  \frac{\partial \mathcal Q^{(1)\alpha}_{\varepsilon+\omega, \varepsilon}}{\partial \Delta^{(1)}}
  = \mathcal{C}^\alpha_{\varepsilon+\omega,\varepsilon} \left(\tau_1 - Q^\alpha_{\varepsilon+\omega} \tau_1 Q^\alpha_{\varepsilon}
     \right),
\ee
\begin{multline}
\label{eqn:app:dQ1/dDelta-K}
  \frac{\partial \mathcal Q^{(1)\text{K}}_{\varepsilon+\omega, \varepsilon}}{\partial \Delta^{(1)}}
  =
  \frac{\partial \mathcal Q^{(1)\text{R}}_{\varepsilon+\omega, \varepsilon}}{\partial \Delta^{(1)}}
  F_{\varepsilon} 
  - 
  F_{\varepsilon+\omega} 
  \frac{\partial \mathcal Q^{(1)\text{A}}_{\varepsilon+\omega, \varepsilon}}{\partial \Delta^{(1)}} 
\\
  + \mathcal{D}_{\varepsilon+\omega,\varepsilon} 
  \big[ \tau_1 - (F_{\varepsilon+\omega}-F_{\varepsilon}) Q_{\varepsilon+\omega}^\text{R} \tau_1 Q_{\varepsilon}^\text{A} \big] .
\end{multline}
\end{subequations}
The first line in Eqs.~\eqref{eqn:app:dQ1/da1Explicit-K} and \eqref{eqn:app:dQ1/dDelta-K} accounts for the modification of the retarded and advanced Green's function, while the second line describes the change in the distribution function.

Our expressions \eqref{eqn:app:dQ1/da1Explicit} and \eqref{eqn:app:dQ1/dDelta} agree with a direct derivation based on the Usadel and self-consistency equations. Notably, they reproduce several known limiting cases: Eq.~(8) of Ref.\ \cite{Ovchinnikov2} (with $\omega=0$), Eqs.\ (32) and (33) of Ref.\ \cite{Moor2017} (with $\bA_0=0$), and Eqs.\ (26) and (29) of Ref.\ \cite{Li2024} (with $\bA_0=0$ and depairing due to magnetic impurities with the spin-flip rate $\Gamma$).

\section{Fluctuation propagator near $T_c$}

In the vicinity of the critical temperature, $\Gamma \ll \Delta$ for all currents up to $I_c$ [see Eq.~\eqref{eqn:5:GammaC}]. Therefore, the polarization operators in Eq.\ \eqref{eqn:2:propSum} can be computed perturbatively in $\Gamma$. In this approximation, we get for the diffuson contribution \eqref{Pid-def}:
\begin{equation}
\label{Pid-Tc}
    \Pi_d(\omega) = - \frac{\pi i }{8 } \frac{\omega}{T} \left( 
    1 -  i \sqrt{\frac{4\Delta^2}{(\omega+i\gamma)^2} - 1}
    \right) ,
\end{equation}
where we neglected the $\omega\,O(\Gamma )$ and $O(T^{-3})$ terms.
To ensure convergency in the ultra-violet, we combine the cooperon contribution $\Pi_c(\omega)$ defined in Eq.~\eqref{Pic-def} with the term $1/\lambda$ and express the latter through the self-consistency equation \eqref{SCE}. The resulting expression can be computed by the deformation of the integration contour over $\varepsilon$ into the upper half-plane, leading to
\begin{equation} \label{eqn:appC:PiCAsymp}
 \Pi_c(\omega) - \frac{1}{\lambda}  = \frac{\pi i}{8} \frac{\omega}{T} - \frac{7 \zeta(3)}{16\pi^2} \frac{4\Delta^2+\omega(\omega+4 i \Gamma)}{T^2}, 
\end{equation}
where we neglected the $O(T^{-3})$ terms. Note that the first terms in Eqs.\ \eqref{Pid-Tc} and \eqref{eqn:appC:PiCAsymp} mutually cancel each other.

In the region of large frequencies, $\omega \gg \gamma$, we neglect $\gamma$ in Eq.~\eqref{Pid-Tc} and $\Gamma$ in Eq.~\eqref{eqn:appC:PiCAsymp}, and get the fluctuation propagator in the form
\begin{equation}
  \frac{1}{L^\text{R}(\omega)}
  =
  \frac{\pi i}{8} \frac{\sqrt{\omega^2-4\Delta^2}}{T}  
  - \frac{7\zeta(3)}{16\pi^2} \frac{4\Delta^2+\omega^2}{T^2}  . \label{eqn:2:fluctPropTTC}
\end{equation}
Note that the last term in the equation smear the peak in $\chi_{\mathrm{SH}}(\omega)$ at $2\Delta$.

In the region of small frequencies, $\omega \ll \Delta$, we neglect $\omega$ in the last term of Eq.~\eqref{eqn:appC:PiCAsymp} and rewrite the term with $7\zeta(3)$ using Eqs.~\eqref{DeltaGL}, \eqref{eqn:5:DeltaGammaGL} and \eqref{eqn:5:GammaC}. Thus we arrive at
\begin{equation}
\label{LR-res-app}
  \frac{1}{L^\text{R}(\omega)}
  = -\frac{\pi \Delta}{4 T} \frac{\omega}{\omega + i \gamma}  - \frac{\pi}{2} \frac{3\Gamma_c - \Gamma}{T}.
\end{equation}
Here we retain only the leading terms in $\Gamma$, neglecting $\omega \, O(\Gamma)$ corrections [see Eq.~\eqref{eqn:appC:PiCAsymp}].

\section{Blocks $\mathcal{I}_d$ and $\mathcal{J}_d$ at $\omega\ll\omega_*$ and $T\to T_c$
\label{app:Asymp}}

\begin{figure}
\centering
\includegraphics[width=.99\linewidth]{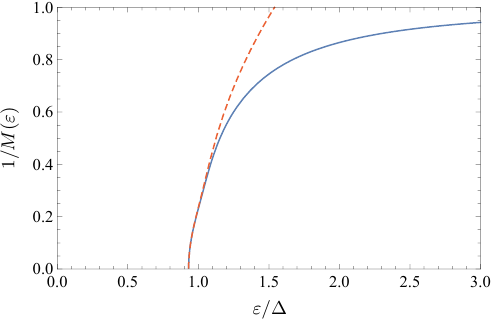}
\caption{The function $1/M(\eps)$, which defines the energy dependence of  the diffuson in Eq.\ \eqref{dif-M}, for $\Gamma/\Delta=0.01$: exact (solid line) and approximate solution \eqref{M-approx} (dashed line).}
\label{F:M}
\end{figure}

Here we calculate the RA blocks $\mathcal{I}_d(\omega)$ and $\mathcal{J}_d(\omega)$ in the overlapping peaks regime, $\omega\ll\omega_*$, near the critical temperature, see Sec. \ref{SSS:overlapping}. 
Since the diffuson denominator at $\omega=\gamma=0$ vanishes above the gap, in the limit $(\omega,\gamma)\ll\omega_*$ it is sufficient to expand $\mathcal{D}_{\eps_+,\eps_-}^{-1}$ to the linear term in $\omega+i\gamma$, leading to Eq.\ \eqref{dif-M} for $\mathcal{D}_{\eps_+,\eps_-}$, where the real function $M(\eps)$ is given by
\begin{align}
\label{M-gen}
  \!
  M(\eps)
  = {} & {}
    (\cos\theta^\text{R}_\eps+\cos\theta^\text{A}_\eps)/2
\nonumber \\{} &
  - (i\Gamma/4)
    [ \sin(2\theta^\text{R}_\eps) - 2\sin(\theta^\text{R}_\eps+\theta^\text{A}_\eps) ]
  \partial_\eps\theta^\text{R}_\eps
\nonumber \\{} &
  + (i\Gamma/4) 
    [ \sin(2\theta^\text{A}_\eps) - 2\sin(\theta^\text{R}_\eps+\theta^\text{A}_\eps) ]
  \partial_\eps\theta^\text{A}_\eps .
\end{align}
Differentiating the Usadel equation \eqref{eqn:3:usadel} and using it once again to eliminate $\eps$, we find
\be
\label{theta'}
  \partial_\eps\theta^\text{R}_\eps
  =
  \frac{i\sin^2\theta^\text{R}_\eps}{\Delta-\Gamma\sin^3\theta^\text{R}_\eps} .
\ee
The function $1/M(\eps)$ calculated numerically with the help of Eqs.\ \eqref{M-gen} and \eqref{theta'} at $\Gamma/\Delta=0.01$ is shown in Fig.~\ref{F:M}. At large energies well above the gap, $M(\eps)\to1$, and Eq.\ \eqref{dif-M} turns to the usual zero-dimensional normal-state diffuson. The function $M(\eps)$ increases with decreasing $\eps$ and diverges at $\eps=E_g$, when the denominator in Eq.\ \eqref{theta'} goes to zero.

As the integrals in Eqs.\ \eqref{eqn:3:Idef} and \eqref{eqn:3:jdDef} are determined by small vicinities of the minigap, we parametrize the energy as 
\be
\label{D1}
  \eps =  ( \Delta^{2/3} + s \Gamma^{2/3})^{3/2} .
\ee
With such a definition, $E_g$ corresponds to $s=-1$, the density of states is peaked around $s=0$, and the integrals for $\mathcal{I}_d(\omega)$ and $\mathcal{J}_d(\omega)$ mainly originate from the region $s\lesssim1$. Right at the minigap, the spectral angle is given by $\theta^\text{R}(E_g)=\pi/2+i\psi_g$, where $\cosh\psi_g=(\Delta/\Gamma)^{1/3}\gg1$.
Explicitly separating this parametrically large contribution, we present $\theta^\text{R}$ in the vicinity of $E_g$ in the form 
\be
\label{D2}
  \theta^\text{R}(\eps) = \chi + i \arccosh[(\Delta/\Gamma)^{1/3}] .
\ee
Substituting Eqs.\ \eqref{D1} and \eqref{D2} into the Usadel equation~\eqref{eqn:3:usadel} and keeping the leading order in $\Gamma/\Delta\ll1$, we obtain a $\Gamma$-independent relation $e^{3i\chi} - 3se^{i\chi} - 2i = 0$. This cubic equation is solved using the Cardano formula, with the physical solution given by
\be
  e^{i \chi}
  = 
  \begin{cases}
    \frac{e^{\pi i/6} \left( 1 + \sqrt{1+s^3 }\right)^{2/3} + e^{-\pi i/6} s}{\left( 1 + \sqrt{1+s^3 }\right)^{1/3}}, 
    & s>-1,
    \\
    2 i \sqrt{|s|} \sin\frac{\pi-\arccot\sqrt{|1+s^3|}}{3},
    & s<-1. 
  \end{cases}
\label{appC:eqn:chiS}
\ee
The advanced angle is obtained as $\theta^\text{A}(\eps)=-[\theta^\text{R}(\eps)]^*$.

Now we substitute Eq.\ \eqref{D2} into Eqs.\ \eqref{M-gen} and \eqref{theta'}, and keep the leading order in $\Gamma/\Delta\ll1$. Thereby we obtain
\be
\label{M-approx}
  M(\eps)
  = 
  \frac{2}{\sqrt{3}}
  \frac{\Gamma^{1/3}}{\Delta^{1/3}}
  \frac{(1+s')^{1/3}}{s+(1+s')^{2/3}} ,
\ee
where we have introduced $s'=\sqrt{1+s^3}$.
The inverse of this function is plotted in Fig.\ \ref{F:M} by the dashed line. We see that it fairly well approximates the exact behavior in the relevant peak region above the minigap.

Performing the same analysis for the Green functions in Eqs.\ \eqref{eqn:3:Idef} and \eqref{eqn:3:jdDef}, we arrive at
\begin{align}
\label{Idw-app}
  \mathcal{I}_d(\omega)
  & = 
  \frac{3^{3/2} i \Delta^{4/3}}{16 \Gamma^{1/3} (\omega+i \gamma)} 
  \int_{-1}^\infty \frac{ds}{s} \, \mathcal{P}(s) ,
\\{}
\label{Jdw-app}
  \mathcal{J}_d(\omega)
  & = 
  -
  \frac{3^{3/2} i \omega}{4\pi (\omega + i \gamma)} \frac{\Delta}{T} 
    \int_{-1}^{\infty} \frac{ds}{s^2} \, \mathcal{R}(s) ,
\end{align}
where
\begin{gather}
  \mathcal{P}(s)
  =
  (1+8s'-9s'^2) t^2 + s^2 (9s'-1) t - 4 s s' ,
\\{}
  \mathcal{R}(s)
  =
  (1+2s'-3s'^2) t^2 + s^2 (3s'-1) t ,
\end{gather}
and $t=(1+s')^{1/3}$.
The integrals in Eqs.\ \eqref{Idw-app} and \eqref{Jdw-app} are given by
\begin{gather}
  \int_{-1}^\infty \frac{ds}{s} \, \mathcal{P}(s) 
  =
  - 
  \frac{2^{11/3} \pi^2}{5\, \Gamma^3(2/3)} ,
\\{}
  \int_{-1}^{\infty} \frac{ds}{s^2} \, \mathcal{R}(s) 
  = 
  \frac{8\pi}{3^{3/2}} ,
\end{gather}
and we arrive at Eqs.\ \eqref{Id-overlap-res} and \eqref{Jd-overlap-res} of the main text.
The same result has been obtained by Ovchinnikov and Isaakyan \cite{Ovchinnikov2}, although their computational approach was not specified.

\section{Superfluid density at $T=0$}
\label{A:SF0}

At absolute zero, equilibrium properties of a superconductor can be determined analytically for an arbitrary supercurrent up to the critical value $I\le I_c$ \cite{Maki1963,Skalski1964}. The key simplification offered at $F(\eps)=\sign\eps$ is the possibility to reduce the energy integral to an integral over the spectral angle $\theta^\text{R}_\eps$, with $\theta^\text{R}_0=\pi/2$, $\theta^\text{R}_{\infty}=0$, and the Jacobian given by Eq.~\eqref{theta'}. The self-consistency equation~\eqref{SCE} thus leads to an implicit relation
\be
\label{eqn:appE:SelfConT0}
  \Delta = e^{-\pi \eta/4} \Delta_0(0) ,
\ee
where $\eta=\Gamma/\Delta$. The general expression for the dc current,
\be \label{eqn:appE:currGen}
  I = \sqrt{2 e^2 D \Gamma} \, \nu 
  \int d\eps \, F(\eps) \im \sin^2 \theta^\text{R}_\eps ,
\ee
then translates to 
\be
  \frac{I}{I_c}
  =
  e^{-3\pi (\eta-\eta_c)/8} 
  \sqrt{\frac{\eta}{\eta_c}} 
  \frac{1 - 4 \eta/3\pi}{1 - 4 \eta_c/3\pi} .
\ee
The critical current, corresponding to $\partial I/\partial \eta = 0$, is achieved at 
\be
  \eta_c = 3\pi/8+2/\pi - \sqrt{(3\pi/8+2/\pi)^2-1} \approx 0.3003 .
\ee

Using the same technique, we now compute the conductivity at $T=0$ in the quasistatic limit $\omega\to0$. Due to the factor $F_{\eps_+}-F_{\eps_-}$, the RA blocks do not contribute to $\Omega_s$, and only the RR-block terms ($K^\text{R}$, $\mathcal{J}_c$, $\mathcal{I}_c$, and $\Pi_c$) require evaluation. Retaining the leading terms in $\omega$ and integrating over the spectral angle $\theta^\text{R}_{\varepsilon}$, we obtain the contributions to $\Omega_s$ from each of the three conductivity terms:
\begin{subequations}
\begin{gather}
  \Omega_1/\pi \Delta_0(0) = 1 - 4 \eta/3\pi ,
\\
  \Omega_2^\text{qp}/\pi \Delta_0(0) = - 8 \eta/3\pi ,
\\
  \Omega_2^\text{SH}/\pi \Delta_0(0) = - 2\pi \eta/(4-\pi \eta) ,
\end{gather}
\end{subequations}
which coincides with Eq.~(46) of Ref.~\cite{Ovchinnikov1}.
A negative sign for the anisotropic contribution to the superfluid stiffness $\Omega_2$ was recently reported in Ref.~\cite{Wang2025}.

For the parallel configuration ($\bA_1\parallel\bA_0$) in the quasistatic regime, the measure of the superfluid density,
\be
\label{OmegasT0qst}
  \Omega_s
  = 
  \Omega_1 + \Omega_2^\text{qp} + \Omega_2^\text{SH} ,
\ee
is shown in the inset to Fig.~\ref{fig:coeffAAtZeroOmega}. It is positive for $I<I_c$, vanishing at the critical current.
At $I=I_c(0)$, $\Omega_2^\text{qp}/\Omega_1 = -0.29$ and $\Omega_2^\text{SH}/\Omega_1 = -0.71$.

In the dynamic regime, $\omega\gtrsim\gamma$, the singular diffuson $\mathcal{D}_{\eps_+,\eps_-}\propto1/\omega$ compensates the smallness of $F_{\eps_+}-F_{\eps_-}$ for $\mathcal{J}_c$, $\mathcal{I}_c$, and $\Pi_c$, and these RA blocks do contribute to the low-frequency behavior of $\im\sigma(\omega)$.

\end{document}